\documentclass[superscriptaddress,groupedaddress,nofootnoteinbib,11pt]{article}
\usepackage{jheppub}

\usepackage{amsmath, amsfonts, amsthm, amssymb, graphicx, hyperref, mathrsfs}
\usepackage[dvipsnames]{xcolor}
\usepackage{subfigure}
\usepackage[utf8]{inputenc}

\allowdisplaybreaks[3]

\def\({\left(}
\def\){\right)}
\def\[{\left[}
\def\]{\right]}
\def\d{\mathrm{d}}

\newcommand{\p} {\partial}
\newcommand{\f}[2]{\frac{#1}{#2}}
\def \bal#1\eal  {\begin{align} #1 \end{align}}
\newcommand{\eref}[1]{{Eq.~(\ref{#1})}}
\newcommand{\be} {\begin{equation}}
\newcommand{\ee} {\end{equation}}
\newcommand{\bc}{\begin{center}}
\newcommand{\ec}{\end{center}}
\newcommand{\bim} {\begin{itemize}[noitemsep]}

\newcommand{\eim} {\end{itemize}}

\newcommand{\nd} {\nabla}
\newcommand{\pd} {\partial}
\newcommand{\mc} {\mathcal}

      \newcommand{\bfx} {{\bf x}}

\newcommand{\ai}{{\alpha}}

\newcommand{\li}{{\lambda}}
\newcommand{\ti}{{\tau}}

\newcommand{\oi}{\omega}

\newcommand{\thi}{\theta}

\title{Excited oscillons: cascading levels and higher multipoles}

\author[a]{Yi-Jie Wang,}
\author[a]{Qi-Xin Xie,}
\author[a,b]{and Shuang-Yong Zhou}

\affiliation[a]{Interdisciplinary Center for Theoretical Study, University of Science and Technology of China,\\ Hefei, Anhui 230026, China}
\affiliation[b]{Peng Huanwu Center for Fundamental Theory, Hefei, Anhui 230026, China}

\emailAdd{yjwang@mail.ustc.edu.cn}
\emailAdd{xqx2018@mail.ustc.edu.cn}
\emailAdd{zhoushy@ustc.edu.cn}
\preprint{\small USTC-ICTS/PCFT-22-28}

\date{\today}

\abstract{

Two types of excited oscillons are investigated. We first focus on spherical symmetry and find that there are a tower of spherical oscillons with higher energies. Despite having multiple approximate ``nodes'' in their energy density profiles, these oscillons are long-lived. We find that during the lifetime of a highly excited oscillon it will cascade down all the lower energy levels before its disintegration. We also point out the existence of excited oscillons with higher approximate multipoles, which generally have shorter lifespans than the spherical ones. Apart from performing nonlinear simulations with absorbing boundary conditions, we also apply a perturbative method to analyze some features of these excited oscillons. 

}

\begin{document}
\maketitle
\flushbottom

\section{Introduction}

While topological defects such as sine-Gordon kinks are easy to find in 1+1 dimensions, Derrick's theorem dictates that (smooth) localized {\it static} solutions are bound to be unstable for a canonical scalar field with a potential in more than 1+1 dimensions \cite{derrick1964comments}. Stable localized solutions do exist in higher dimensions, in the form of Q-balls \cite{Friedberg:1976me,Coleman:1985ki,Lee:1991ax}, if the scalar is endowed with an internal symmetry and one settles to {\it stationary} solutions that are time periodic. Nevertheless, even in the case of a real scalar field, there are some localized objects that are like Q-balls but quasi-periodic and live for a long time, nowadays known as oscillons \cite{Bogolyubsky:1976yu,hep-ph/9503217}. 

Oscillons commonly exist in models that admit potentials with attractive self-interactions, which translates to some relatively mild conditions on the flatness of the potential, similar to the conditions for the Q-balls to exist \cite{Coleman:1985ki,Lee:1991ax}. The properties of oscillons have been extensively studied over the years; see, {\it e.g.}, \cite{hep-ph/0110065, hep-th/0203072, hep-th/0505273, hep-th/0606016, hep-th/0609023, hep-th/0610191, hep-th/0610267, 0712.0614, 0802.3525, 0804.0791, 0903.0953, 0910.5922, 1003.3459, 1201.1934,  1210.2227,  1303.1102, 1306.3868, 1401.6168,  1504.04038, 1508.01028, 1612.07750,  1708.01344, 1708.08922, 1803.08550,  1809.07724, 1901.06130,   1907.00611, 1908.11103, 1910.04128, 1911.03352, 1911.03340,  2010.05933, 2010.07789,  2012.13409, 2104.02069, 2105.01089}. Indeed, the current cosmological observations indicate that the inflaton potential may be very flat \cite{Planck:2018jri}, and oscillons can be copiously generated in the reheating period after inflation or in other similar processes in the early universe \cite{hep-ph/0503081, hep-th/0602187, hep-th/0604134, 0712.3034, 1002.3380, 1006.3075, 1009.2505, 1103.1911, 1106.3335, 1511.02336, 1710.06851, 1711.10496, 1912.09658, 2012.14697}. The presence of oscillons can affect the Big Bang thermal history and lead to an oscillon dominated epoch. The production of oscillons in the early universe is often accompanied by stochastic gravitational waves that characterize the energy scales and other features of the underlying model  \cite{1304.6094,1501.01217,1607.01314,1707.09841,1712.03231,1801.03321,1803.08047,1902.06736,1905.00371,1907.10613,2011.12201,2112.07626,2204.07152}. The oscillon preheating scenario has also been investigated in full numerical relativity \cite{1912.09658,2010.05933,2112.07626}, and it is found that primordial black holes can sometimes form from them in the case of strong gravity effects \cite{1912.09658,2010.05933}.  Furthermore, oscillons in 2+1 dimensions have been observed in laboratories \cite{umbanhowar1996localized, lioubashevski1999oscillons, arbell2000temporally}.

In this paper, we will search for and investigate oscillons that are excited in two ``orthogonal'' ways. Firstly, we will focus on excited oscillons in spherically symmetry. We find that there are excited oscillons with increasing levels of energy, whose oscillation frequency decreases with the level of the energy. As the evolution of an oscillon is a process where its (dominant) oscillation frequency adiabatically increases until it reaches the upper limit and dissipates, these different energy plateaus naturally appear in sequence in the evolution history of a highly excited oscillon, forming a rather intriguing cascade (see Figure \ref{fig:energyevolution}). These higher energy oscillons are solutions with increasingly more (approximate) nodes, and the lifespans of these plateaus decrease with their energies. We will also provide a semi-analytic computation of the lifespans of the plateaus by simply approximating the oscillon with a factorizable background plus the leading perturbative radiation field, which matches the nonlinear simulations rather well.

Excited spherical Q-balls with multi-node structures have been studied in Refs \cite{hep-th/0205157,1206.2930, 2004.03446, 2112.00657, 2201.09239}.  The spherical oscillons we studied in this paper are actually different in spirits, apart from the difference between a complex and a real scalar. The equivalent spherical oscillons similar to those excited Q-balls can be also constructed, as we will briefly discuss in the paper, but those strongly multi-node spherical oscillons are much shorter lived than our weakly multi-node ones. This is most easily seen in the perturbative construction where our spherical oscillons can be approximated by an un-excited background plus a leading multi-node perturbation.

Secondly, we will construct non-spherical oscillons that are excited to have higher multipoles. These oscillons may arise from anisotropic environments or from collisions of oscillons, perhaps similar to the fact that astrophysical black holes generally have spin. In any rate, these are nonlinear localized objects with interesting properties that are worth exploring. 
Non-spherical Q-balls, the counterparts for the case of a complex scalar field, have been recently studied in the forms of charge-swapping Q-balls \cite{1409.3232, 2101.06988, 2202.08392} or in the forms of spinning Q-balls and their generalizations in the presence of gauge fields or strong gravitational effects \cite{hep-th/0205157, hep-th/0302032, gr-qc/0505143, 0804.1357, 0812.3968, 0904.4802, 0907.2801, 0907.0913, 0909.2505, 1207.3715, 1612.05835, 1909.01950}. Whereas one can construct a Q-ball with a fixed multipole, this is impossible for an oscillon, which nevertheless, as we will show, can have a dominant multipole supported by some weak subdominant ones. Therefore, unlike a 2+1D Q-ball whose profile is factorizable into a radial part plus a rotation phase, a multipolar oscillon is generally more complex in nature. 

To numerically construct a spinning oscillon, one can take a factorized configuration as the input, and we see that it can then relax to the spinning solution. We find that it is imperative to prepare the initial configuration to at least solve the background equation of motion. This suggests that the ``attractor basin'' for spinning oscillons are relatively small, compared to the spherical case, where one can obtain excited oscillons from quite generic configurations. Also, for higher multipolar oscillons, we have only found spinning oscillons but not the swapping ones. This is due to the fact that for the case of oscillons the real scalar field only has one component, while for the case of Q-balls the complex field has two components. Because of this, the latter can be intuitively viewed as two inter-connected oscillons, which allows one to arrange attracting, opposite-charge lumps to form charge-swapping Q-balls \cite{1409.3232}. These results show that the similarities between oscillons and Q-balls are also shared for their composite/excited structures, with some major caveats as well. 

The paper is organized as follows. In Section \ref{sec:modelsetup}, we introduce the model we will focus on in this paper and the perturbative expansion we will use for the semi-analytical computations later. We describe the numerical codes we use for the fully nonlinear simulations in Appendix \ref{sec:Scode} and some technical details about the perturbative expansion in Appendix \ref{sec:decouple}. In Section \ref{sec:casosci}, we investigate properties of spherical cascading oscillons including their evolution patterns, lifetimes, insensitivity to initial conditions and spectra. The efficient spherical code used in this section is validated by the full-blown Cartesian code in Appendix \ref{sec:sph3Dsim}. We also perform a perturbative analysis to compute the lifespans of the various energy plateaus of the cascading oscillons. In Section \ref{sec:multipolarOSC}, we investigate another type of excited oscillons, which have higher multipoles. The simulations in this section are performed with a Cartesian code with absorbing boundary conditions. We conclude in Section \ref{sec:conlusions}.

\section{Model and setup}

\label{sec:modelsetup}

We will focus on oscillons from the simplest model with only one real scalar field $\varphi$ that is invariant under reflection $\varphi \to -\varphi$. As we will see, this simple model already gives rise to rich time evolutions and complex structures that are quasi-stationary. From the modern, prevailing point of view of effective field theory, such a theory in $d+1$ dimensions is given by the action
\bal
S_{\rm t}=\int \d^{d+1}\tilde{x}\bigg( & -\frac{1}{2}\partial_{\tilde \mu}\varphi \partial^{\tilde \mu}\varphi- \frac{1}{2}m^2\varphi^2+\lambda\varphi^4- {g}_0 \varphi^6  +  g_1 \varphi^8  + g_2 (\pd_{\tilde \mu}\varphi\pd^{\tilde \mu} \varphi)^2 +\cdots
\bigg),
\eal
where $m$ is the mass of the scalar and the mass dimensions of the coupling constants are $[\li]=3-d,~[g_0]=4-2d,~[g_1]=5-3d,~[g_2]=-d-1$ and so on. (One may add terms like $\pd^4\varphi^2$ and $\pd^2\varphi^4$, but these terms can be removed by field re-definitions.) For simplicity, we will in this paper only consider the case of $d=2$, and also truncate to the order of $\varphi^6$ in the Lagrangian:
\be
S=\int \d^{3}\tilde{x}\bigg( -\frac{1}{2}\partial_{\tilde \mu}\varphi \partial^{\tilde \mu}\varphi- \frac{1}{2}m^2\varphi^2+\lambda\varphi^4- {g}_0\varphi^6 \bigg) .
\ee
In 2+1 dimensions, the $\varphi^4$ term is relevant with $\li$ having mass dimension $[\li]=1$ and the $\varphi^6$ term is marginal, so $S$ is apparently renormalizable. Making use of dimensionless variables 
\be
x^\mu=m\tilde{x}^\mu, ~~~\phi=\f{(2\lambda)^{1/2}\varphi}{m},~~~ g=\f{ m^2{g}_0}{2\lambda^2} ,
\ee
the action can be re-written as
\begin{equation}
S=\f{m}{2\li}\int \d^{3}x\;{\cal L} \equiv \f{m}{2\li}\int \d^{3}x\(-\frac{1}{2}\partial_\mu\phi \partial^\mu\phi-\frac{1}{2}\phi^2+\frac{1}{2}\phi^4-\frac{1}{2}g\phi^6\).
\label{action}
\end{equation}	
This leaves us with just one dimensionless free parameter $g$ in the model, and we shall explore the dependence of the phenomena on this parameter. Results for models with different hierarchies between $m$ and $\li$ can be extracted by appropriate scalings. For the simple case with no hierarchy between $m$ and $\li$ (say, $m=2\li$), one can view the values of all the dimensionful quantities in this paper as in units of the mass. The (re-scaled) energy density Legendre-transformed from $\mc{L}$ is given by
\begin{equation}
{\cal H}=\frac{1}{2}\dot{\phi}^2+\frac{1}{2}(\nd\phi)^2+\frac{1}{2}\phi^2-\frac{1}{2}\phi^4+\frac{1}{2}g\phi^6,
\label{ham}
\end{equation}
where the dot stands for partial derivative with respect to $t$, and the equation of motion is given by
\begin{equation}
\ddot\phi- \nabla^2 \phi +\phi -2\phi^3+3g\phi^5=0.
\label{overallpde}
\end{equation}

As discussed in the introduction, action (\ref{action}) supports quasi-stable oscillons, which are nonlinear, localized quasi-solitons that exist in many field theories having shallow potentials. Unlike a Q-ball solution that exists in a complex scalar theory, a spherically symmetric oscillon for a real scalar does not simply factorize to a spatial profile and a temporal oscillation. Rather, the oscillon solution Fourier-decomposes into a spectrum of multiple modes, although, as we shall see later, many of its interesting properties can be described by the dominant one, with the higher frequency modes determining the decay rate of the oscillon. We shall also see that there exist a series of excited spherical oscillons, which naturally emerge as different phases a highly excited oscillon goes through during its entire lifetime. In addition to spherically symmetric oscillons, in Section \ref{sec:multipolarOSC}, we shall later see that there are also complex oscillons that contain higher multipoles and may be viewed as the real scalar counterparts of charge-swapping Q-balls \cite{1409.3232, 2101.06988, 2202.08392} and spinning Q-balls \cite{hep-th/0205157, hep-th/0302032, gr-qc/0505143, 0804.1357, 0812.3968, 0904.4802, 0907.2801, 0907.0913, 0909.2505, 1207.3715, 1612.05835, 1909.01950}. We shall later construct and simulate these solutions fully nonlinearly.

The simulation schemes we use in this paper are explained in Appendix \ref{sec:Scode}. In Section \ref{sec:casosci}, where we focus on spherically symmetric oscillons and perform multiple parameter scans, we will use a spherical grid code, which is more efficient as it is essentially a 1+1D code and whose absorbing boundary conditions are easier to implement. A spherical code, however, neglects the effects of non-spherical perturbations. In Appendix \ref{sec:sph3Dsim}, we compare a typical evolution in the spherical code with that in the Cartesian code, and show that the effects of non-spherical perturbations are negligible for our case. On other hand, all the simulations in Section \ref{sec:multipolarOSC} are run with the Cartesian code, as we will deal with higher multipolar configurations there.

To help construct the complex, excited oscillons and understand their properties semi-analytically, we will use the following mode expansion that can capture many salient features of these solutions
\begin{equation}
\phi(t,r,\theta)=\Phi_{l_0}(r) r^{l_0} \cos(\omega t-l_0\theta) + \sum_{n>1,|l|>l_0}A_n^l(r) r^{|l|} \cos(n\omega t-l\theta).	
\label{modeexp}
\end{equation}
where $l_0,l,n$ are integers (with $n>1$ and $l_0\ge0$)\,\footnote{Negative $l_0$ is of course allowed, representing an opposite spinning background, but we restrict $l_0$ to be non-negative for simplicity since the theory is parity invariant.} and we have factored out $r^l$ such that regularity at $r=0$ requires that 
\be
\pd_r\Phi_{l_0}(r=0)=0,~~~\pd_r A_n^{l}(r=0)=0. 
\ee
This expansion for the $l_0=0$ case has been previously adopted in \cite{2004.01202} for investigating the lowest energy spherical oscillons. As assumed in mode expansion (\ref{modeexp}) that will be verified later, an oscillon solution usually has a dominant oscillation frequency $\oi$, whose amplitude $\Phi_{l_0}(r)$ can be viewed as the {\it background} amplitude, and the {\it perturbations} around this factorizable background have frequencies that are mostly multiples of the base frequency. For a higher multipolar oscillon ($l_0\neq 0$), as will be explained later, the background mode needs to be spinning in one direction such as  $\cos(\omega t-l_0\theta)$.
By examining the equations of motion explicitly (see Appendix \ref{sec:decouple}), one can also show that when the background is spinning in one direction, the opposite spinning perturbative modes are decoupled (and unexcited), so in \eref{modeexp} we can restrict $l$ to be $l>l_0$ in the summation without loss of generality. 

For later convenience, we write the nonlinear terms in the equations of motion as $F(\phi)\equiv -2\phi^3+3g\phi^5$, and define $C_n^{l}(r)$ and $B_n^{l}(r)$ as the coefficient functions in the following expansions
\bal
\label{defC}
F\(\Phi_{l_0} r^{l_0} \cos(\omega t-l_0\theta)\)&=\sum_{i\ge 0}C_{2i+1}^{(2i+1) l_0}(r)\cos((2i+1)\omega t-(2i+1) l_0\theta),
\\
\label{defB}
F'\(\Phi_{l_0} r^{l_0} \cos(\omega t-l_0\theta)\)&=\sum_{i\ge 0}B_{2i}^{2i l_0}(r)\cos(2i\omega t-2i l_0\theta).
\eal
For example, we have $C_1^{l_0}=-\frac{3}{2}r^{3l_0}(\Phi_{l_0})^3+\frac{15}{8}gr^{5l_0}(\Phi_{l_0})^5$ and $B_0^{0}=-3r^{2l_0}(\Phi_{l_0})^2+\frac{45}{8}gr^{4l_0}(\Phi_{l_0})^4$. Then, the background equation of motion is given by
 \begin{equation}
  \partial_{r}^2\Phi_{l_0}+\frac{2l_0+1}{r}\partial_r\Phi_{l_0}+(\omega^2-1)\Phi_{l_0}-\frac{C_1^{l_0}}{r^{l_0}}  =0 ,
  \label{Phiequs}
\end{equation}
and the perturbative equations of motion are given by
\be
\label{Aequs}
\partial_{r}^2A_n^l+\frac{2l+1}{r}\partial_r A_n^l+(n^2\omega^2-1-B^0_0)A_n^l -\frac{C_n^l}{r^l}
-\frac{1}{2r^l}\(\sum_{j=2}^{\infty}B_j^{jl_0}r^{l+jl_0}A_{n+j}^{l+jl_0}+\sum_{j=2}^{n-2}B_j^{jl_0}r^{l-jl_0}A_{n-j}^{l-jl_0}\)=0.
\ee
The presence of only a single angular momentum mode in the background field implies that many of the perturbative fields are actually un-excited. From the definitions of $C_n^l$ and $B_n^l$ (\eref{defC} and \eref{defB}) above, we see that $C_{2i}^l=0$ and $B_{2i+1}^l=0$, so the equations of motion for $A_{2i}^l$ are decoupled from those of $A_{2i+1}^l$ and are unsourced ({\it i.e.,} without the inhomogeneous term in their equations of motion), leading to $A_{2i}^l=0$. Additionally, all the $A^{nl_0+k}_n$ modes with nonzero $k$ are unsourced, and are thus also unexcited. (The equations of motion for $A^{nl_0+k}_n$ with different $k$ are decoupled from each other, and only the  $A^{nl_0}_n$ modes are sourced; See Appendix \ref{sec:decouple} for more details.) 

The background field $\Phi_{l_0}$ and perturbative fields $A_n^l$ can be solved with standard ODE solvers. As $\Phi_{l_0}$ falls off to zero very fast asymptotically at large $r$, going like $e^{-\sqrt{1-\omega^2}r}/r^{({2l_0+1})/{2}}$, the background equation (\ref{Phiequs}) can be easily solved by a standard shooting method, shooting from near $r=0$ to a large $r$. On the other hand, from the perturbative equations of motion (\ref{Aequs}), we see that the radiation field $A_n^l$ falls off to zero much slower and oscillatorily at large $r$, going like $H_b(r)\propto J_{l}(\sqrt{n^2\omega^2-1}r)/r^{l}\sim \cos(\sqrt{n^2\omega^2-1}r -(2l+1)\pi/4)/r^{l+1/2}$, where $J_l$ is the Bessel function of the first kind. (It is also possible to have a solution that includes the Bessel function of the second kind, but that only affects the solution by a constant phase in the cosine, which does not affect the radiation rate we are after in this paper.) Therefore, to accurately solve \eref{Aequs}, we use a shooting procedure where we shoot from a small $r$ to a relative large $r=r_b$ and match the value of ${A_n^l(r_b)}/{\partial_r A_n^l(r_b)}$ to the value of ${H(r_b)}/{\partial_r H(r_b)}$ at $r_b$.

\section{Cascading oscillons}

\label{sec:casosci}

In this section, we will focus on composite structures in spherically symmetry oscillons. We will see that there are excited oscillons whose energy decays cascadingly with time, forming a series of descending steps, and we explore their properties and explain this phenomena with a semi-analytical mode expansion. In this section, since we are interested in spherically symmetric solutions, we will make use of a spherical code, which is easier to setup the absorbing conditions at the boundaries and much cheaper computationally. This of course has the danger of neglecting non-spherical modes in the evolutions. In Appendix \ref{sec:sph3Dsim}, we validate our approach with full 2+1D simulations and show that the non-spherical modes are negligible in these simulations.

\begin{figure}
	\centering
	\includegraphics[width=9cm]{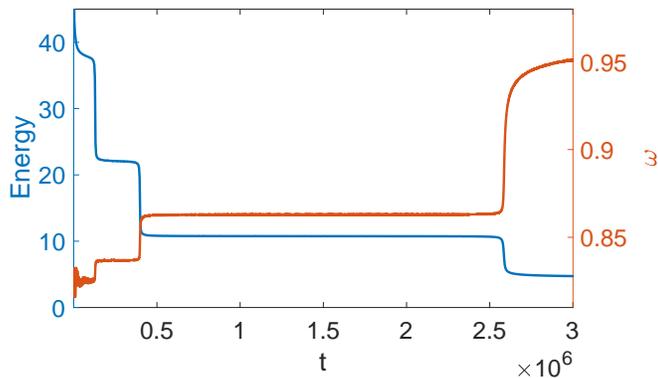}
	\caption{Cascading levels of the energy $E$ (defined in \eref{define:E}) and frequency $\oi$ (defined to be the (angular) frequency of a point in the center of the oscillon) of a spherically symmetric oscillon. We can see four distinct energy plateau levels as the oscillon evolves, and the frequencies of the levels increase as the energy steps down.  The initial configuration is chosen to have $A=1,a=81,\sigma=90$, and the coupling is $g=0.60$. We label the energy levels from the lowest (the 1st level) to the highest. We have not seen the decay of the last energy plateau ({\it i.e.,} the 1st energy level) in our simulation for this renormalizable model in 2+1D.}
	\label{fig:energyevolution}
\end{figure}

\subsection{Cascading levels}

\begin{figure}
\center
\centering
	\subfigure[1st level]{\includegraphics[width=7.7cm]{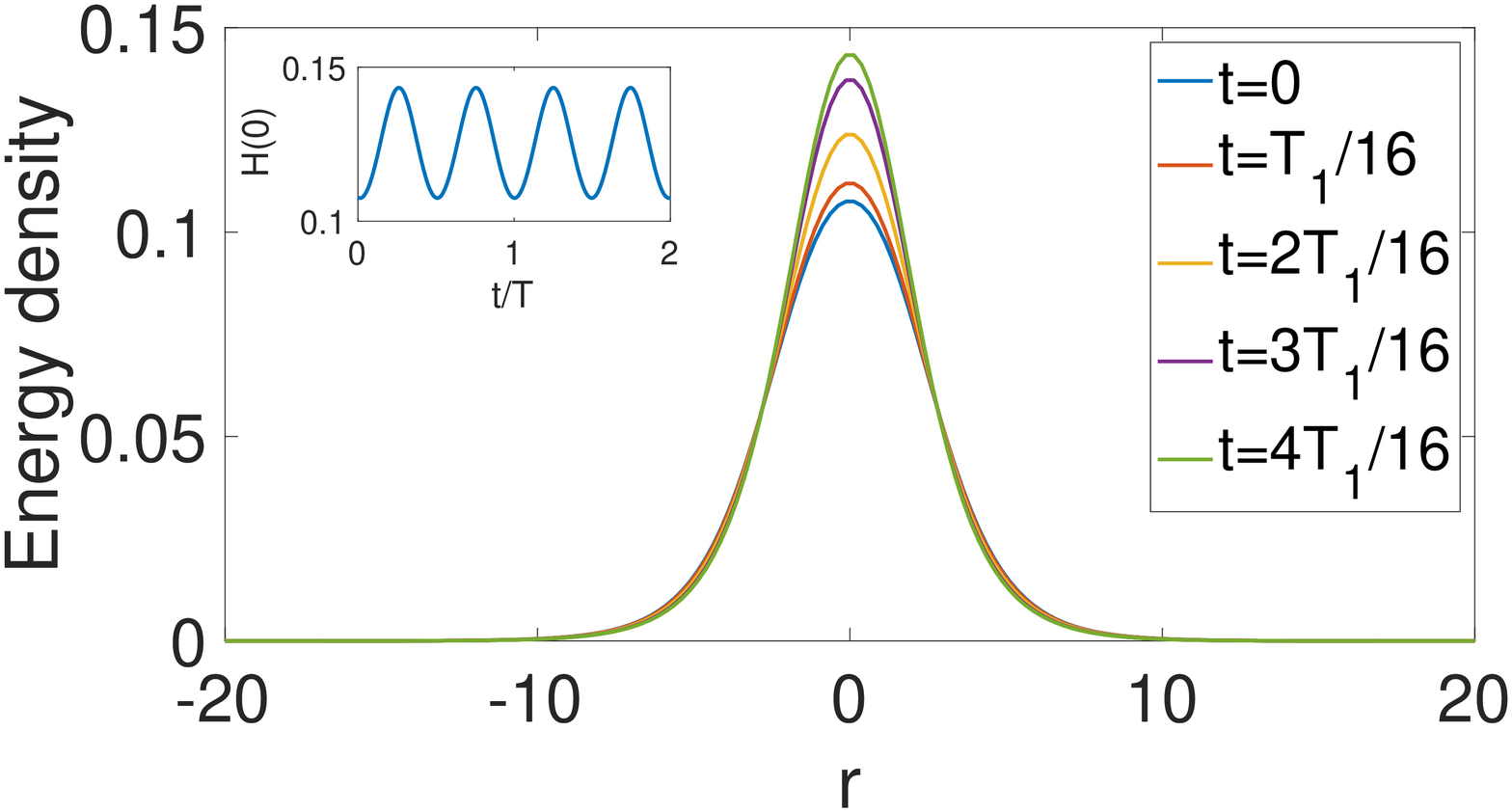}}
	\subfigure[2nd level]{\includegraphics[width=7.7cm]{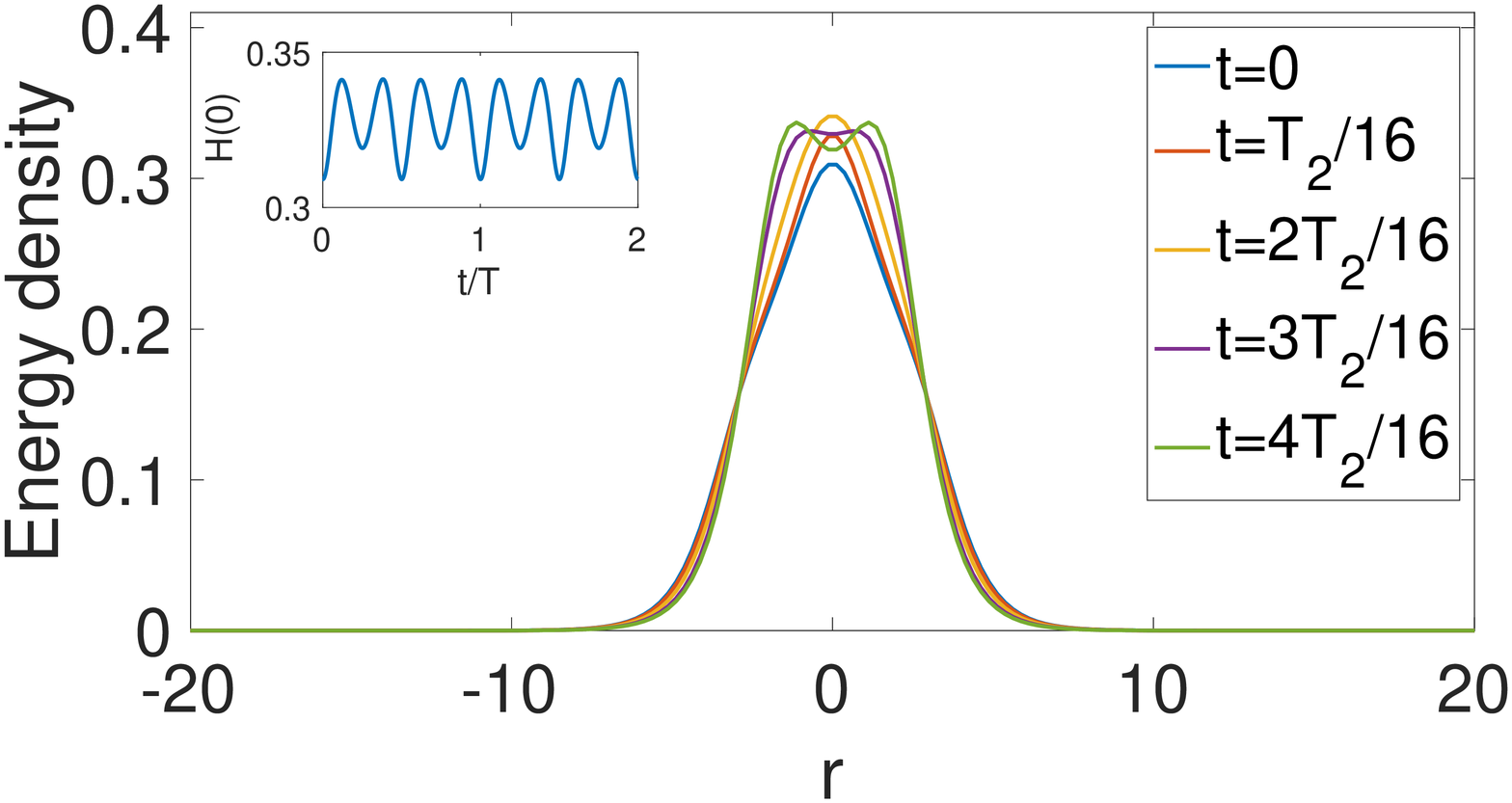}}
	\subfigure[3rd level]{\includegraphics[width=7.7cm]{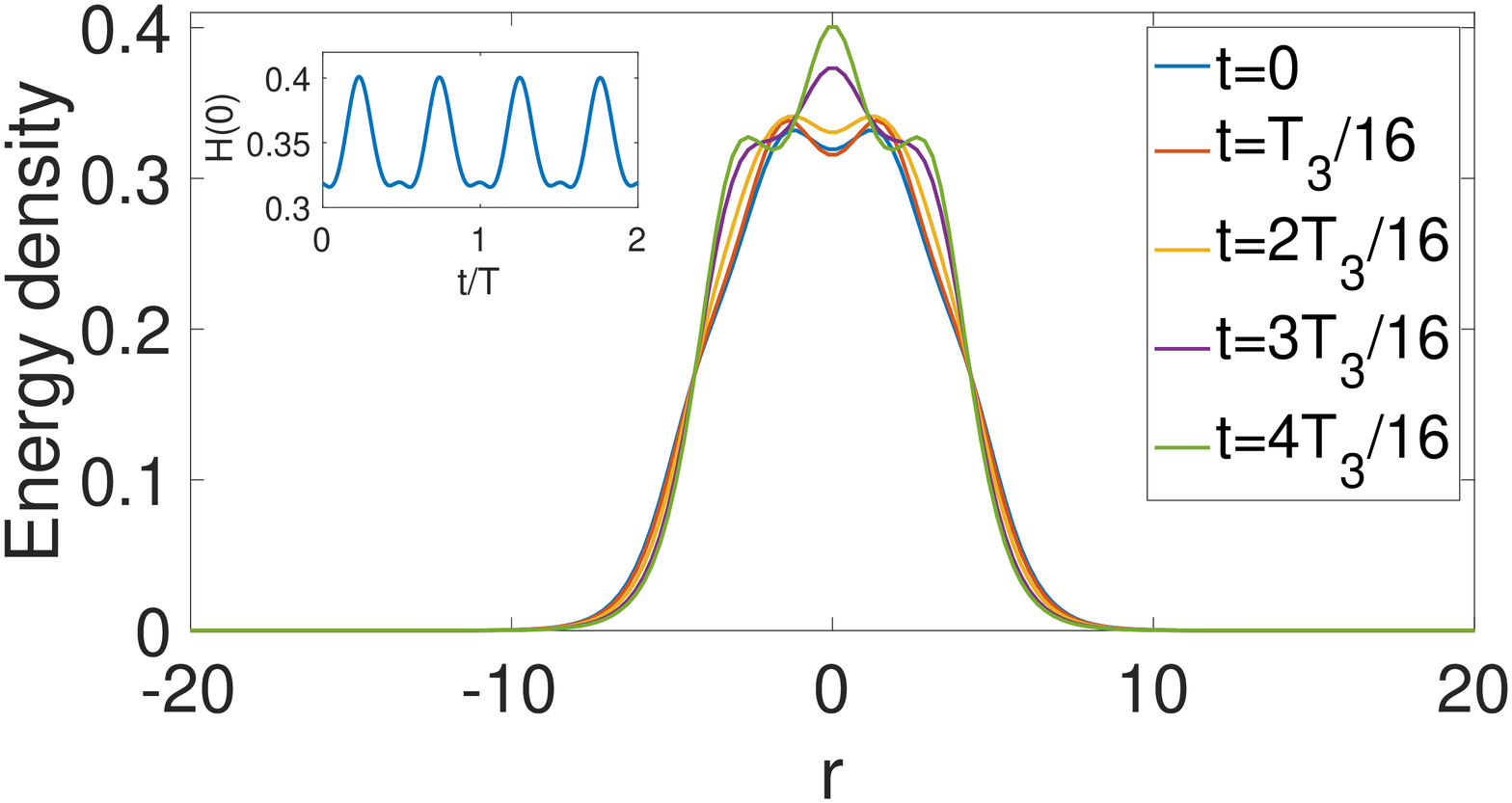}}
	\subfigure[4th level]{\includegraphics[width=7.7cm]{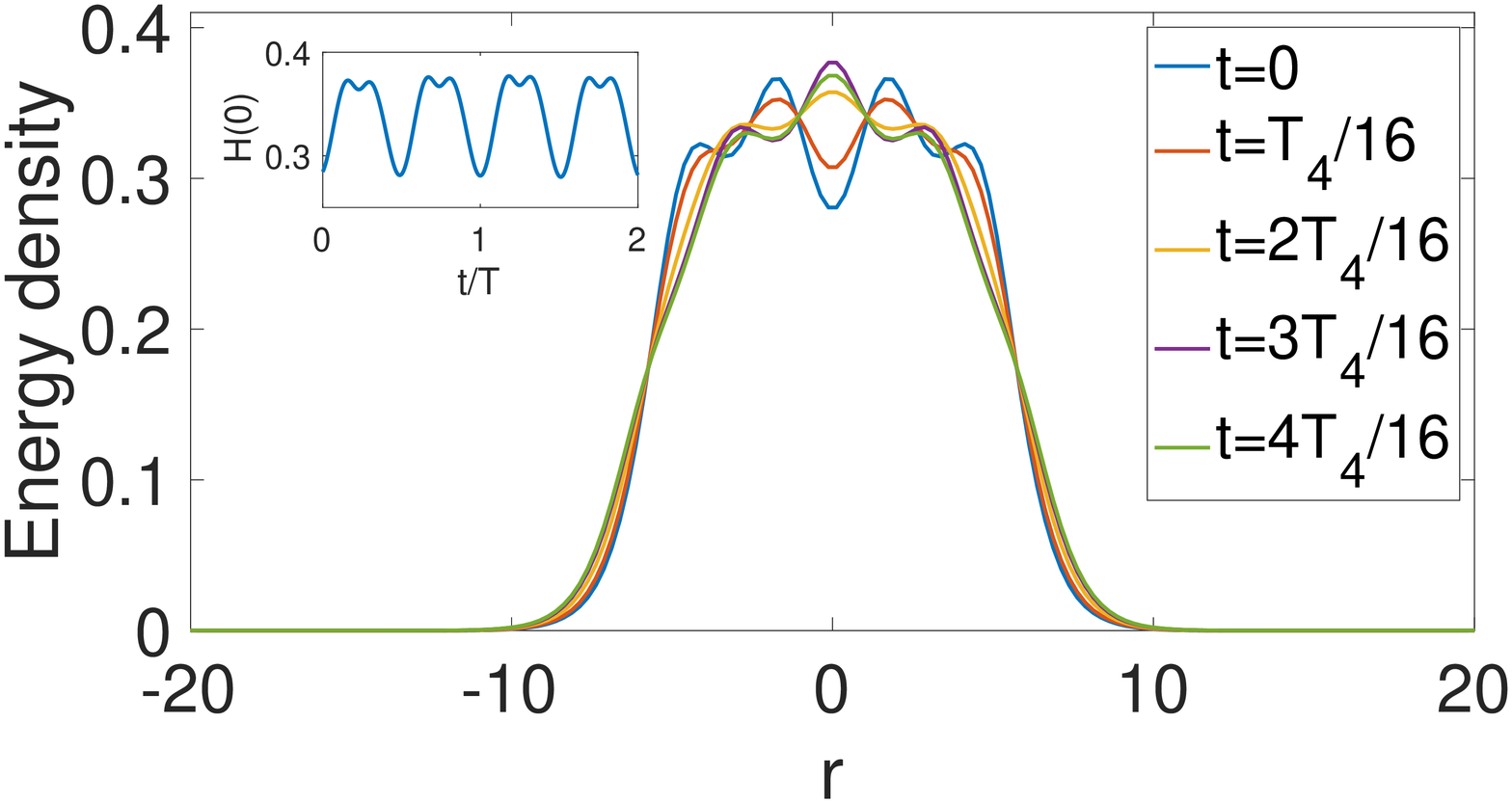}}
	\caption{Evolution of the energy density ${\cal H}(t,r)$ of the 4 plateau levels of the oscillon in Figure \ref{fig:energyevolution}. $T_1, T_2,T_3,T_4$ are the oscillation periods of the field $\phi$ at the center of the oscillon respectively, with $T_1$ being that of the 1st energy level. The oscillation period of the energy density is half of the corresponding period of the field. The insets show the time evolutions of the energy density at $r=0$.}
	\label{fig:Hfig}
\end{figure}

To construct excited spherically symmetric oscillons, we set up a ring-shaped initial configuration (we are working in 2+1D for simplicity):
\begin{equation}
 \phi=A e^{-\frac{(r^2-a)^2}{\sigma^2}},\ \ \dot\phi=0.
\label{initial}
\end{equation}
where $A$ sets the amplitude of the field, $\sigma$ is the thickness of the ring and $a$ determines the radius of the ring. This Gaussian-like, ring-shaped initial setup is of course for convenience, and it will initially relax and shed away some amount of energy as we see in Figure \ref{fig:energyevolution}. When excited oscillons are properly formed, they are more complex than a ring, as we see in Figure \ref{fig:Hfig}. Nevertheless, the relaxed configuration still continuously radiates away energy, and thus the absorbing boundary conditions discussed in Appendix \ref{sec:Scode} are useful to eliminate the unwanted fluctuations in the simulation grid.

Figure \ref{fig:energyevolution} shows a typical evolution of the total energy $E$ (defined in \eref{define:E}) of such an oscillon. Numerically, we shall define the energy of the oscillons by integrating over the energy density for all the points within a radius of $r=20$ from the center of the oscillon:
	\begin{equation}
		E(t)=\int_{r<R} \d^2 x{\cal H}(t,r) .
		\label{define:E}
	\end{equation} 
The energy density of a spherical oscillon typically halves at around $r\simeq 2.5-5$. We can see an interesting phenomenon that the energy $E$  of the oscillon cascades with time, producing multiple levels/plateaus before settling down to the most (quasi-)stable level. We have not seen the sudden decay of the most stable level in our 2+1D simulations for this particularly model, but the oscillon does continuously radiate slowly, which is well-known ({\it e.g.,} \cite{hep-ph/9503217}). As we can see in Figure \ref{fig:energyevolution}, the excited higher energy levels live for much less times, and as will be explained shortly in Figure \ref{fig:Hfig}, they have increasingly more ``nodes'' in their radial profile\,\footnote{Here, the term ``node'' is used in a loose sense, and it is not the point where the field profile remains zero at all times. By a node, we mean a point in the energy density profile that oscillates visually less than the surrounding points; see Figure \ref{fig:Hfig}.}. We can also see that during the evolution within an energy level the dominant (angular) frequency of the oscillon $\omega$ is stable, where frequency $\oi$ is numerically defined as the frequency of the grid point that initially has the largest field amplitude, {\it i.e.}, a point in the center of the oscillon. Also, the level of the energy and frequency matches with each other. The node structure of the different levels follows the pattern that the $n$-th level has $n$ nodes, if we count from $r=0$ to a large positive $r$. For example, in the subfigure (a) of Figure \ref{fig:Hfig} , we see that the radial energy profile has one node at $r\simeq 2.63$, while in the subfigure (d) we have 4 nodes at $r\simeq 1.13,2.63,3.62,5.75$. While some nodes such as the outermost node are not moving in the oscillation, conforming to the exact definition of a node, some other nodes do move slightly and are less well-defined. Remarkably, the peak densities of the excited plateau levels are roughly of a similar value, which is noticeably greater than that of the first level. The increase in energy from the 2nd level to the 4 level is mainly due to the oscillon being fatter for the higher levels. Visually, in Figure \ref{fig:Hfig}, we see that the evolution of the energy density of the excited oscillons have the effects of breathing in and out. This is also technically true for the 1st level oscillon, due to the presence of a node, but is less obvious visually. Incidentally, for the setup of Figure \ref{fig:energyevolution}, the energy levels are: $E\simeq 38, ~22,~11,~6$, which are mostly 2 multiples of the ``base" energy level $E\simeq 6$. For a different coupling constant $g$, the gap between the 1st level and the 2nd level will change accordingly, but the pattern of the gaps between the higher levels seem to remain. This presumably is due to the fact that the higher energy levels come from the increase of the number of nodes in the solution, {\it i.e.,} the energy density profile becoming fatter, with the factor 2 coming from the fact that the spatial dimension is 2.

\begin{figure}
	\centering
	\includegraphics[width=8cm]{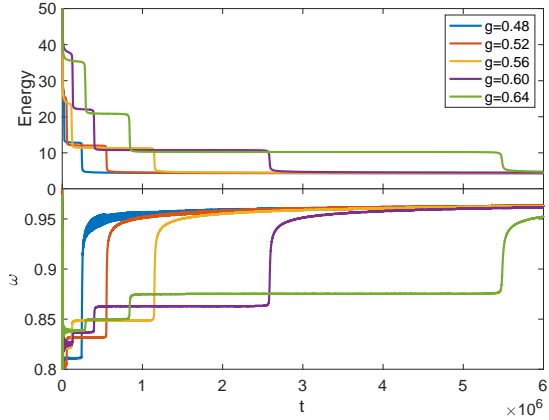}
	\caption{Dependence of the energy levels on the coupling $g$. The simulation setup is the same as Figure \ref{fig:energyevolution} except for different $g$.  While the energy levels of the oscillons are insensitive to the coupling, the lifespans of the levels increase with $g$.  The frequency $\oi$ changes noticeably for different $g$ except for the 1st level, the most stable level.}
	\label{fig:energyevolutionwithg}
\end{figure}

\begin{figure}[tbp]
	\centering
	\subfigure{\includegraphics[width=7cm]{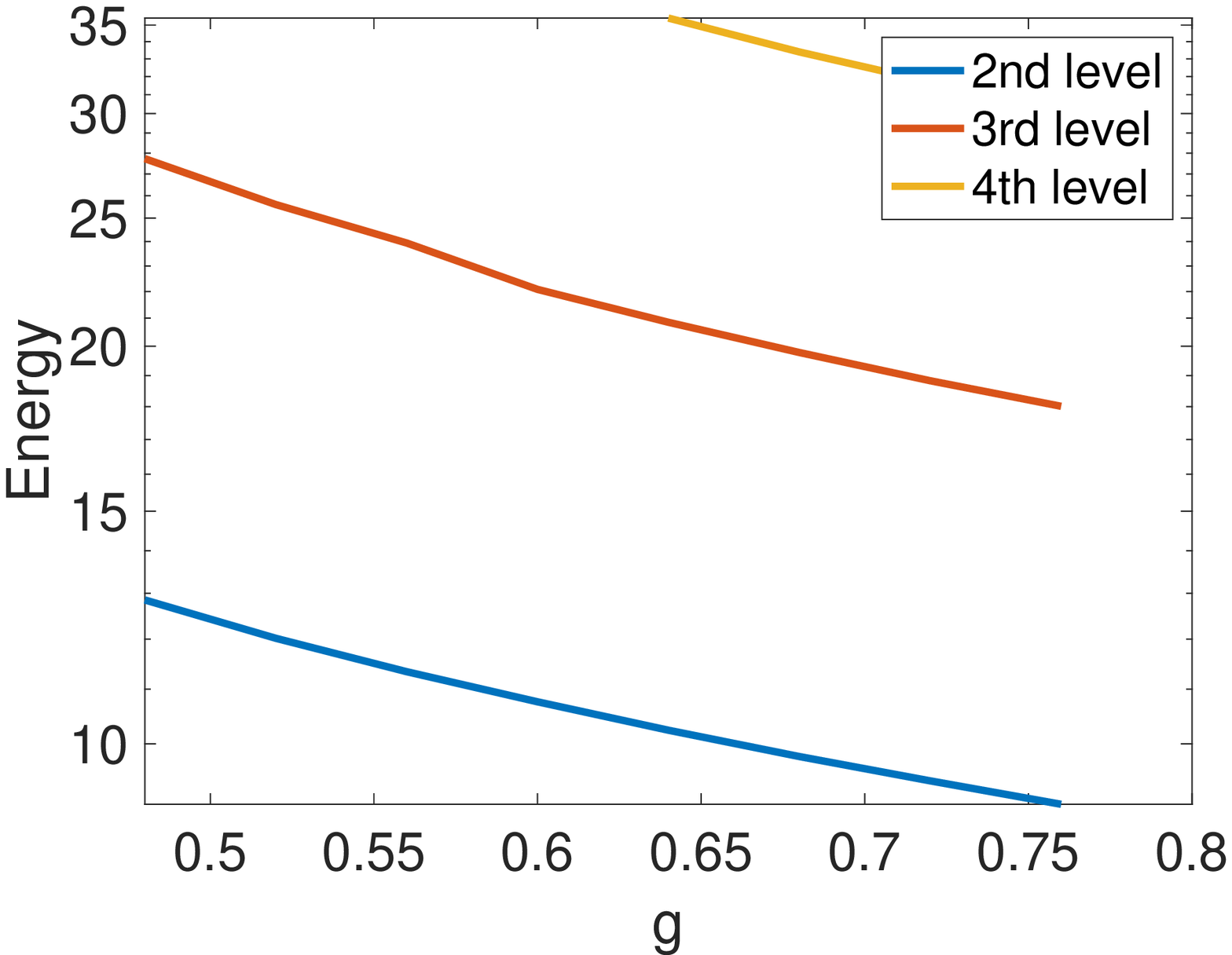}}
	\subfigure{\includegraphics[width=7cm]{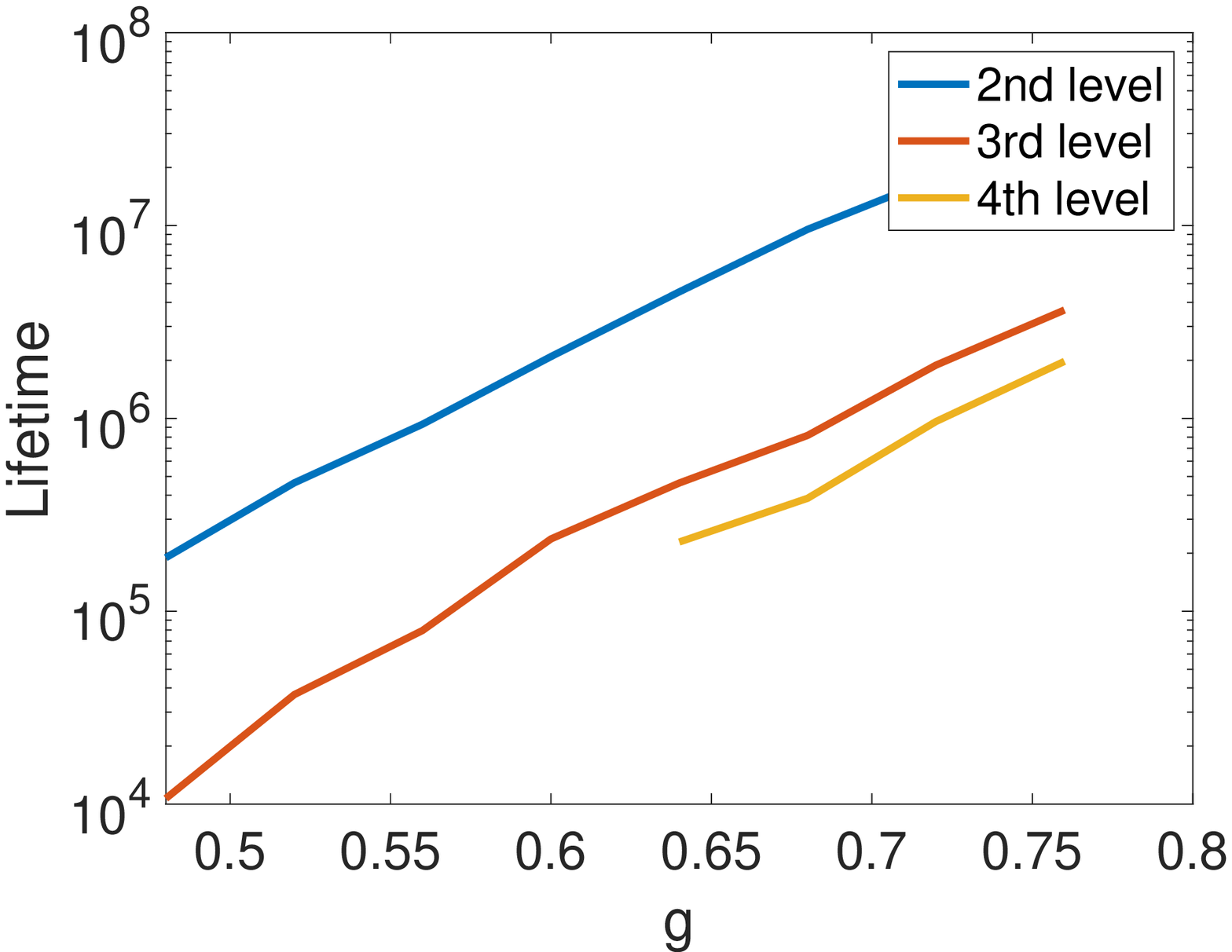}}
	\caption{Scaling behaviors of the energy values (left) and lifetimes (right) of the various levels on the coupling $g$. The $n$-th level of the oscillon refers to the $n$-th lowest energy level. The simulation setup is the same as Figure \ref{fig:energyevolution} except for different $g$.}
	\label{fig:gscaling}
\end{figure}

In our renormalizable model (\ref{action}), there is only one theory parameter, the coupling $g$. (To obtain the behaviors of the solution for different masses and $\varphi^4$ couplings, we can simply scale the units of the solution.) It is instructive to see how the excited oscillons depend on this parameter. In Figure \ref{fig:energyevolutionwithg}, we plot the dependence of the energy levels on the coupling $g$. We see that for a larger coupling the excited energy levels exist for much longer times, but the values of the energy levels only vary slightly for different $g$. Nevertheless, upon careful examination, we find that both the energy values and lifetimes of the levels scale exponentially with $g$; see Figure \ref{fig:gscaling}.  Numerically, we take the lifetime of a plateau of the oscillon $\ti$ to be the difference between the end of the plateau $t_{\rm end}$ and the start of the plateau $t_{\rm start}$:
	\begin{equation}
		\ti_{}=t_{\rm start}-t_{\rm end}. 
		\label{define:lifetime}
	\end{equation} 
For definiteness, $t_{\rm end}$ is taken to be the time when the energy decay rate $P(t)$ is one tenth of $P(t)$ in the middle of the level plateau and the start $t_{\rm start}$ is taken to be ten times of $P(t)$ in the middle of the level plateau, where the energy decay rate can be simply computed by numerically evaluating $P(t)=|E(t+\d t)-E(t)|/{\d t}$.

\begin{figure}[tbp]
 \centering
 \subfigure[]{\includegraphics[height=3.2cm]{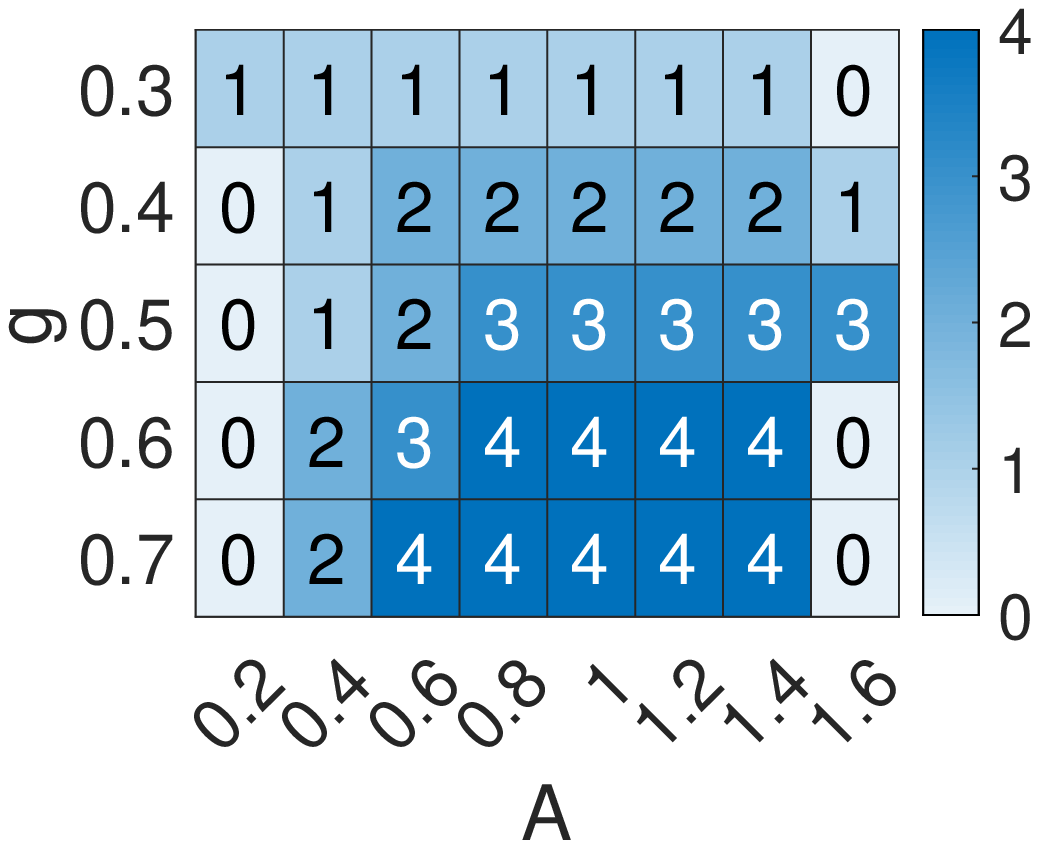}}
 \subfigure[]{\includegraphics[height=3.2cm]{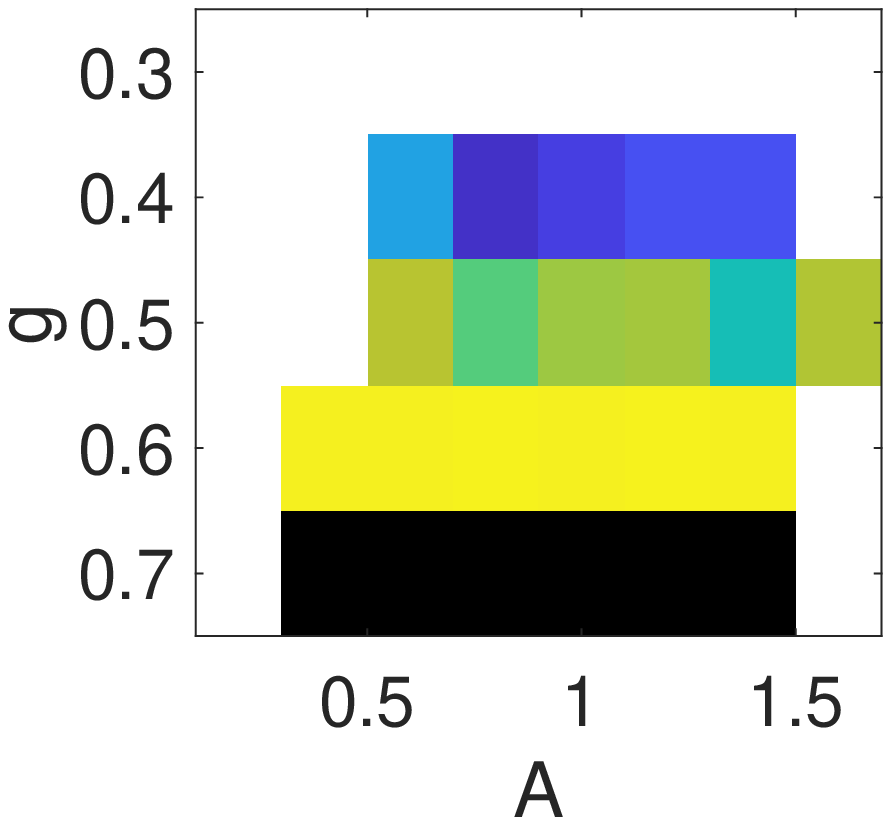}}
 \subfigure[]{\includegraphics[height=3.2cm]{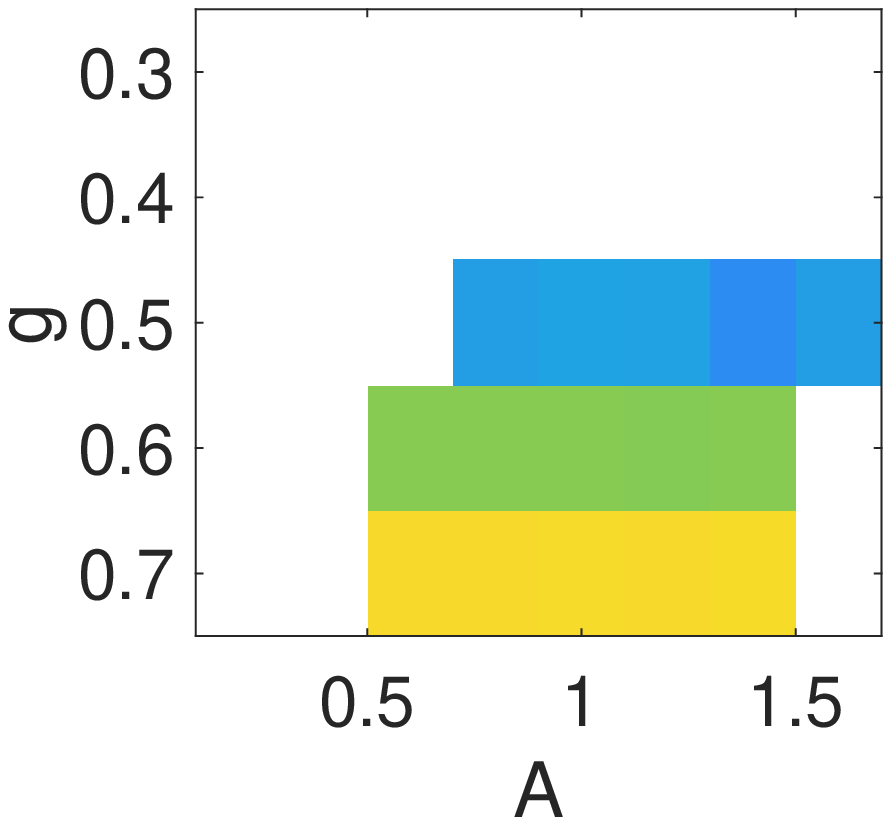}} 
 \subfigure[]{\includegraphics[height=3.2cm]{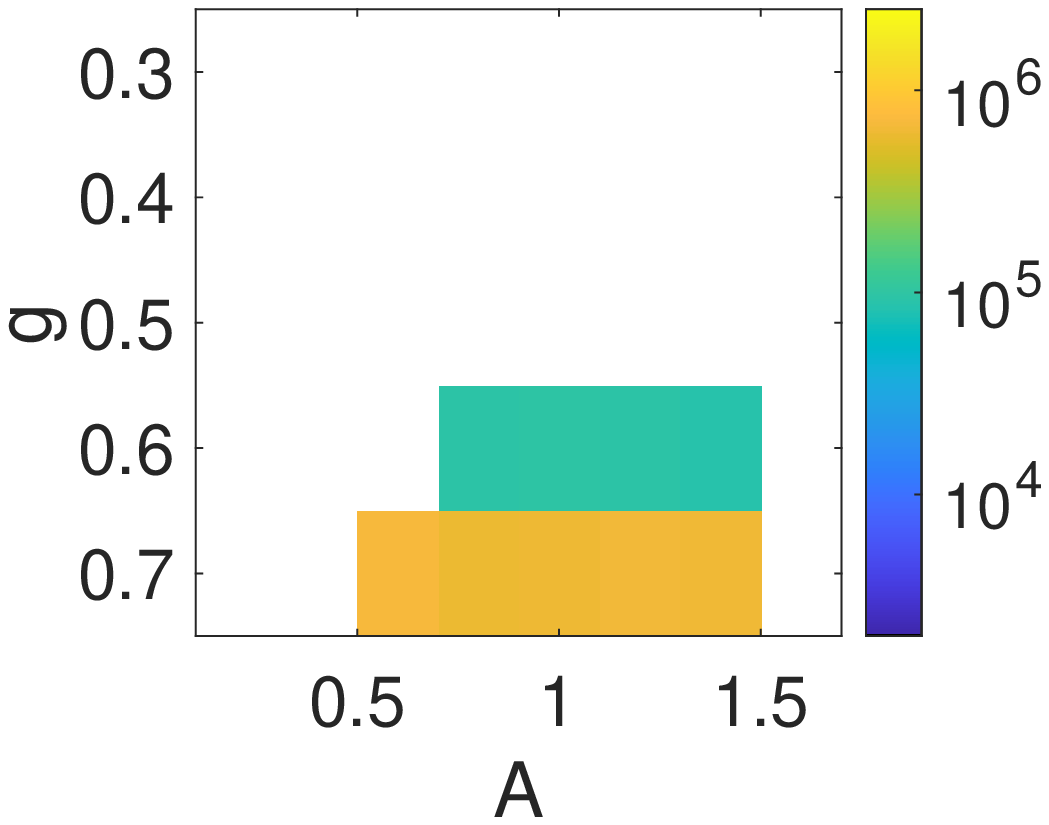}} 
 \caption{Dependence of the lifetime of each plateau on $A$ and $g$. We set $a=81,~\sigma=90$ for a fiducial choice. (a) shows the number of resolvable energy levels for the corresponding $A$ and $g$. (b), (c) and (d) show respectively how the lifetime of the 2nd, 3rd and 4th level depends on $A$ and $g$. The white regions are where the oscillon cannot have the relevant level, while the black regions are the cases where the level has not decayed when time reached $t=5\times10^6$. }
 \label{fig:numandtofplatform}
 \end{figure}

\begin{figure}[tbp]
  \centering
 \subfigure[Energy $E$ of 3rd level]{\includegraphics[width=3.8cm]{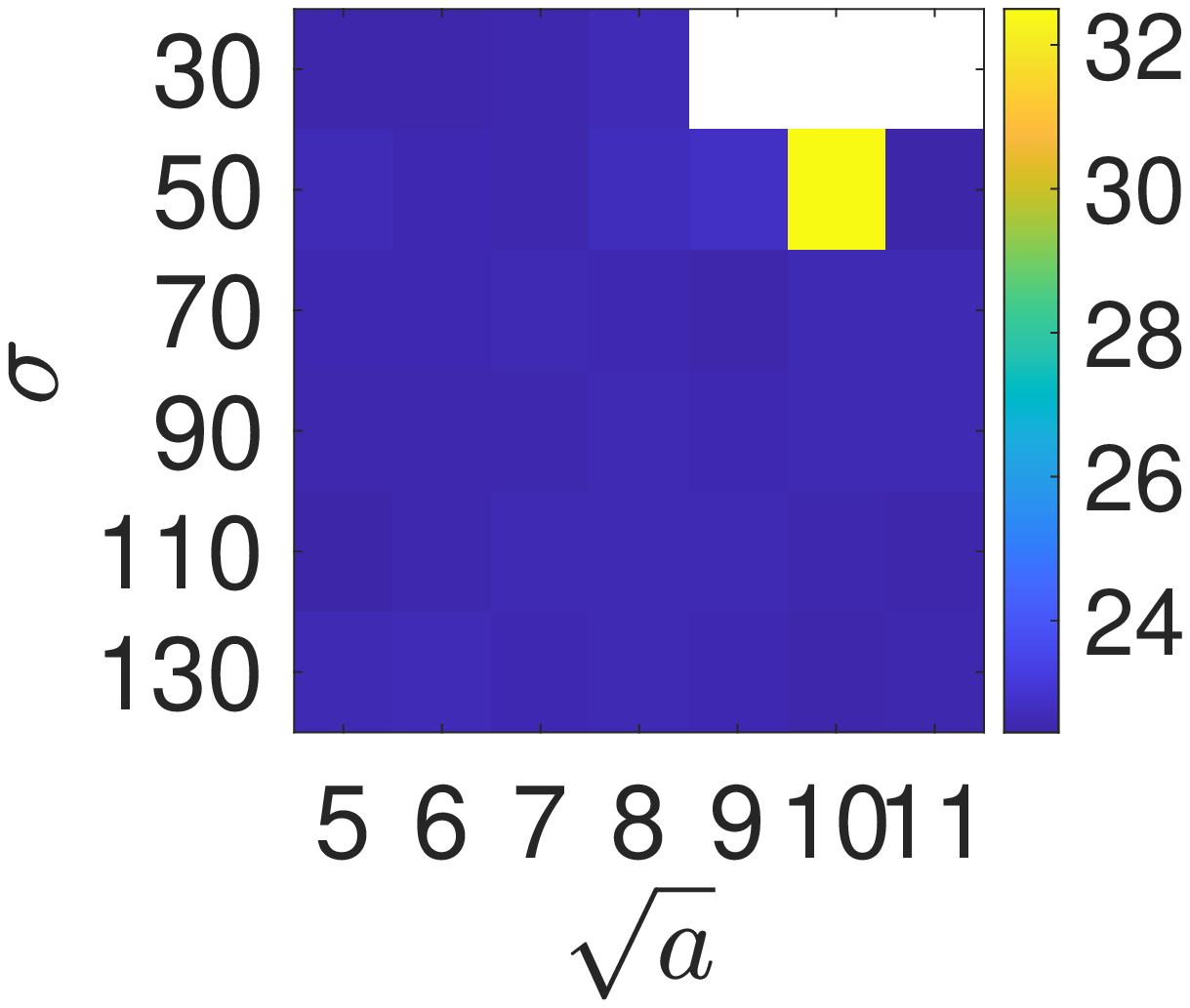}}
 \subfigure[Lifetime of 3rd level]{\includegraphics[width=3.8cm]{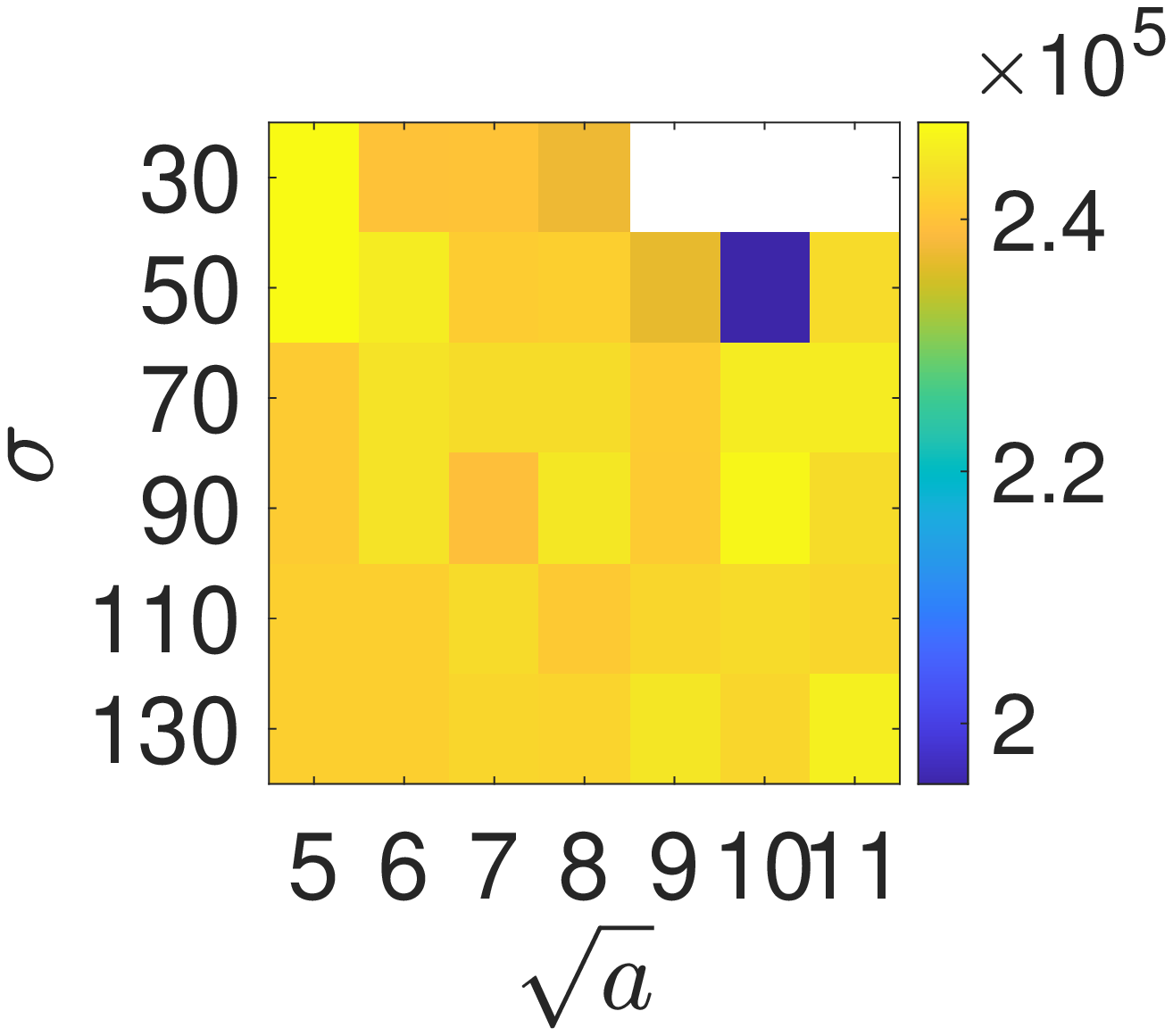}}
 \subfigure[Energy $E$ of 4th level]{\includegraphics[width=3.8cm]{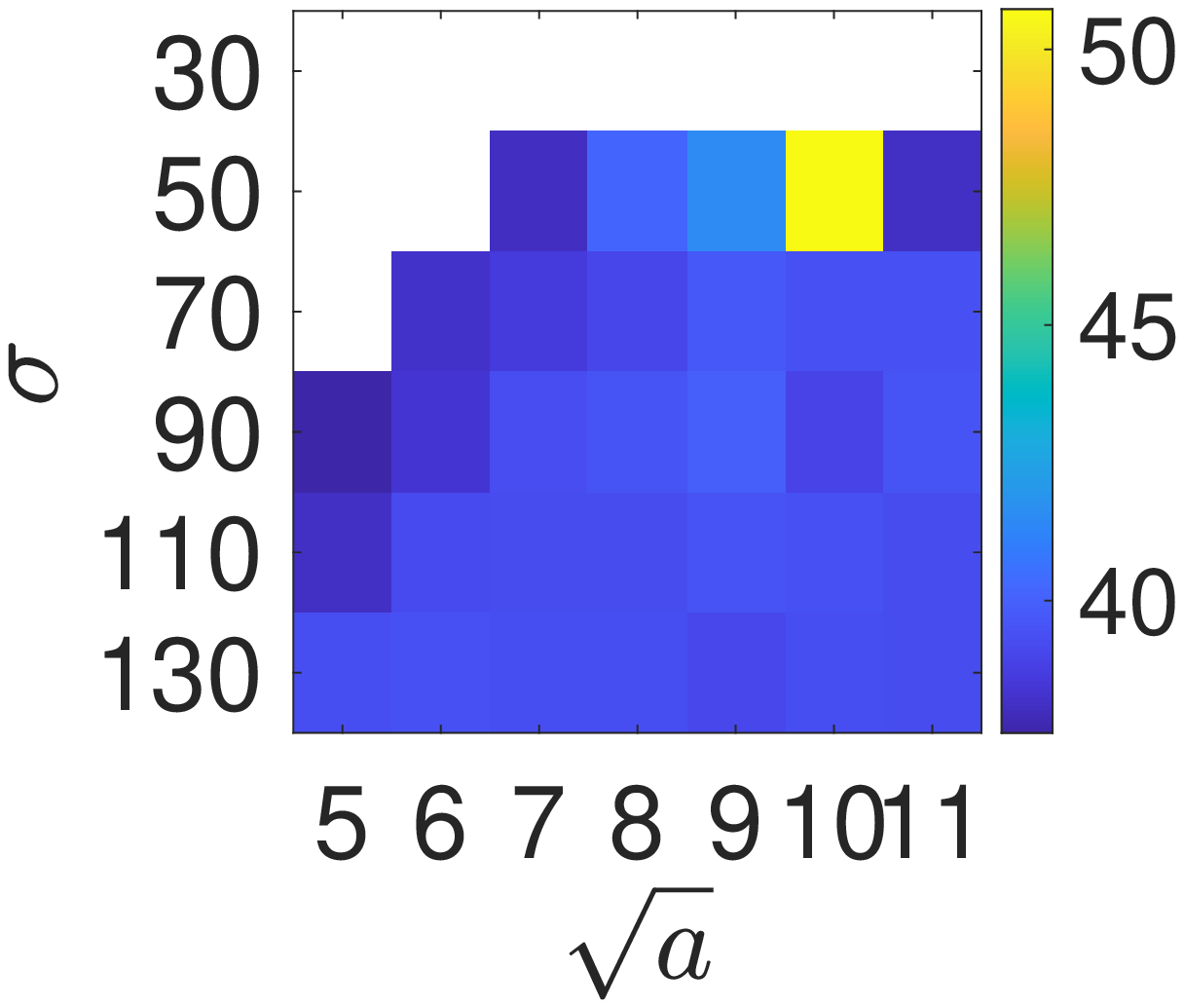}}
 \subfigure[Lifetime of 4th level]{\includegraphics[width=3.8cm]{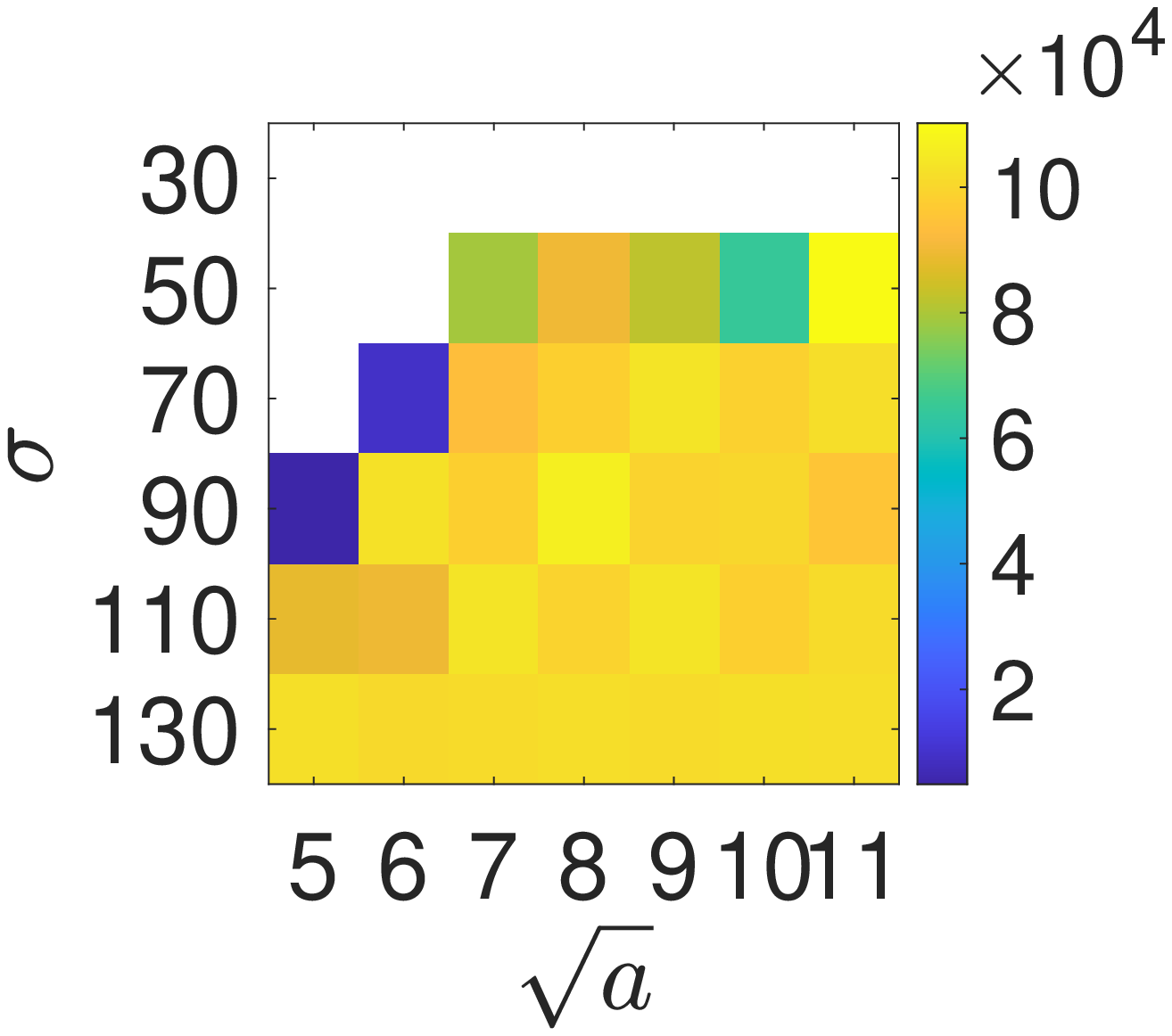}} 
 \caption{Dependence of energies and lifetimes of different levels on initial configurations. $A=1.0$ and $g=0.60$. We see that the energies and lifetimes are not sensitive to the initial $a$ and $\sigma$ except near the boundary of the parameter space where oscillons can form, meaning that these oscillons are some kind of attractors. The white regions are where the oscillon cannot have the relevant level.} 
 \label{fig:initial}
\end{figure}

 In Figure \ref{fig:numandtofplatform}, we plot how the lifetime of each level changes for different $A$ and $g$. As expected, we find that $A$ has little effect on the lifetime as long as the relevant plateau does form in the evolution; the excess of the energy introduced by a larger $A$ is simply shredded away in the initial relaxation.  The lifespans of the displayed energy levels also remain mostly unchanged upon varying other initial conditions. In Figure \ref{fig:initial}, we explore how the energies and lifetimes of the 3rd and 4th levels depend on the values of $a$ and $\sigma$ in \eref{initial}. We see that they are not sensitive to the initial $a$ and $\sigma$ except near the boundary of the parameter space within which oscillons can form. This is a reflection that these energy levels are quasi-stable attractors of the system and justifies our sloppy choice of the initial configuration in \eref{initial}, as the system will evolve to the attractor as long as it is within the attractor basin. For the spherical oscillons, the attractor basin is relatively sizable. It is possible to have higher energy levels should one fine-tunes the initial conditions, which has not been vigorously pursued in this study. However, their lifetimes are much shorter.

\begin{figure}[tbp]
\centering
 \subfigure{\includegraphics[width=8cm]{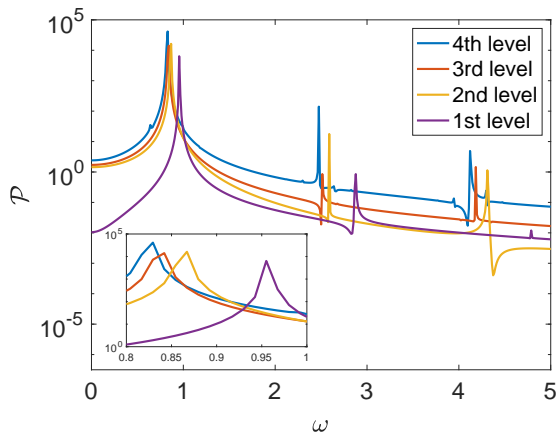}}
 \caption{Fourier spectra ${\cal P}$ of the scalar field at the center the spherical oscillon.  The inset highlights the dominant peak near $\oi=1$. The higher frequency peaks are almost odd multiples of the base frequency near $\oi=1$. The parameters for this plot are $g=0.60$, $A=1.0$, $a=81$ and $\sigma=90$, and the sampling times are $t_a=2.1\times10^4,1.75\times10^5,1.75\times10^6,3.9\times10^6$. }
 \label{fig:fourier}
 \end{figure}

It is also instructive to sample some Fourier spectra of the different energy levels of the oscillon. An example of these is shown in Figure \ref{fig:fourier} where we plot the power spectra of the scalar field at a point in the center of the oscillon:
\be
\mc{P}=|\tilde\phi(\oi,\bfx_{0})|^2,~~~{\rm where ~~} \tilde\phi(\oi,\bfx_{0}) = \int^{t_b}_{t_a} {\d t} e^{-i\oi t} \phi(t,\bfx_{0}) ,
\ee
where $t_a, t_b$ are chosen to be within a flat part for each plateau of the oscillon and $t_b-t_a=500$. It has a distinct peak structure: there is a dominant base frequency, which essentially determines the oscillation period of oscillon, followed by smaller peaks around odd multiples of the base frequency. This justifies our semi-analytical expansion scheme (\ref{modeexp}), which we will use in the next subsection. Turning the argument around, the reason why there are only odd multiples of the base frequency can be seen in the perturbative analysis (\ref{modeexp}); see the text below \eref{Aequs}. Also, we see that the dominant peak frequency decreases as the energy of the level increases so that the most stable level, the 1st level, has the highest dominant peak frequency. The same is also true for the sub-leading peaks. Some kind of structural differences between the most stable level and the higher levels are also noticeable both in the frequencies and the peak powers, with the three higher levels staying more close to each other. This fits in to the picture that the evolution of the oscillon is a process where its base frequency migrates to pass the limit of $\oi=1$, where $\sqrt{1-\omega^2}$ changes from a real value to an imaginary value and the nature of the solution changes from oscillatory to dissipative. For example, the far field changes from $\phi\sim e^{-\sqrt{1-\omega^2}r}/r^{1/2}$ to $\phi\sim \cos(\sqrt{\omega^2-1}r)/r^{1/2}$, the latter allowing an energy outflow. Thus, the different energy plateaus can naturally appear in sequence during the entire lifetime of an excited oscillon.

\subsection{Perturbative analysis}
\label{sec:pertanaly}

In this subsection, we shall use the expansion scheme (\ref{modeexp}) to analyze some of the properties of  the cascading oscillons explored with lattice simulations above. Particularly, we shall semi-analytically compute the lifespans of the energy levels. 

The first order of business in the perturbative analysis using the expansion scheme (\ref{modeexp}) is to nonlinearly solve the background solution $\Phi_{0}(r)$. The approximation of the ``background'' oscillon with only a single mode is a rather crude one, and yet as we will see that at least for our particular model this is quite a good approximation. In fact, to first order approximation, we shall view the oscillon as consisting of this background plus a perturbative radiative field, which is the first excited $A^l_n$ field ($A^0_3$ in this particular case).

\begin{figure}
	\centering
	\includegraphics[width=8cm]{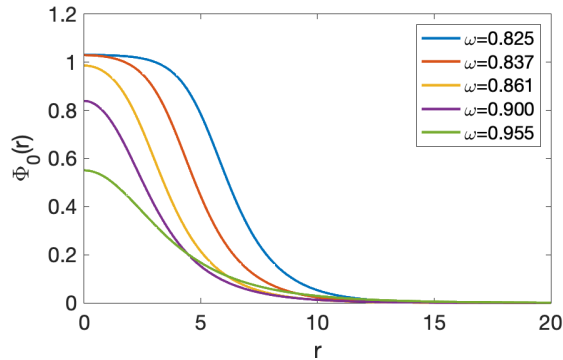}
	\caption{Background profile $\Phi_0(r)$ for different $\omega$. The parameters are the same as those of Figure \ref{fig:energyevolution}.}
	\label{fig:background of l0=0}
\end{figure}

With the expansion scheme (\ref{modeexp}), obtaining the background solution is fairly easy, as it only involves solving the following ordinary differential equation:
 \begin{equation}
  \partial_{r}^2\Phi_{0}+\frac{1}{r}\partial_r\Phi_{0}+(\omega^2-1)\Phi_{0}+\frac{3}{2}(\Phi_{0})^3-\frac{15}{8}g(\Phi_{0})^5 =0.
  \label{back pde}
  \end{equation}
 This can be easily done by a standard shooting method, shooting from near $r=0$ to a large $r$, as mentioned in the last section. To the leading order approximation, we shall simply estimate the energies of the oscillon levels with the background solution 
\be
\phi(t,r)\simeq \Phi_{0}(r) \cos\omega t  .
\ee
To compare with the fully nonlinear solution of the last subsection, we set the initial parameters to be $A=1$, $a=81$ and $\sigma=90$, and choose $g=0.60$ as our fiducial model in this subsection. The $\Phi_{0}$ profiles for a few different $\oi$ can be found in Figure \ref{fig:background of l0=0}.  A comparison between the energies of the oscillons computed from this perturbative approach (``perturbative energy") and from the nonlinear simulations in the last subsection (``nonlinear energy'')  can be found in Table \ref{ntcompare}, showing good agreements between the two.

\begin{table}
\centering
	\begin{tabular}{ | c | c | c | c |}
		\hline
			& 4th	 level & 3rd level & 2nd level \\ 
		\hline
		 perturbative energy 	& $39.67$		& $23.22$		 & $11.75$ \\ 
		\hline
		nonlinear energy	& $\approx39$		& $\approx22$		 & $\approx11$  \\ 
		\hline 
	\end{tabular}
	 \caption{Comparison between the energy computed with only $\Phi_{0}(r)$ and the energy obtained from nonlinear simulations in the last subsection. The $n$-th level refers to the $n$-th energy plateau of the oscillon. The nonlinear energies are only approximate values because the energy does change slightly over the whole lifespan. The parameter setup is the same as Figure \ref{fig:energyevolution}.}
	\label{ntcompare}
\end{table}

With the background established, to determine the lifetimes of the plateau levels, we need to estimate the decay rate or radiation emission rate of the oscillon. To this end, we include the perturbations or radiation fields $A^0_n$ in \eref{modeexp}, which are sourced by the background field. As $A^0_2$ vanishes (cf.~the text below \eref{Aequs}), the lowest order radiation field is $A^0_3$. We shall only take this next leading order approximation. This is retrospectively justified by computing the quintuple contribution, or by simply inspecting the spectra in Figure \ref{fig:fourier} and noticing that the power of the quintuple frequency is far weaker than that of the triple frequency. Therefore, the perturbation equation that needs solving is simply
\begin{equation}
\begin{aligned}
\partial_{r}^2A_3^0+\frac{1}{r}\partial_r A_3^0+(9\omega^2-1-B_0^0)A_3^0 =C_3^0+{\cal O}(A^2).	
\end{aligned}
\label{cal:A30}
\end{equation}
where $B_0^0$ and $C_3^0$ are defined in \eref{defB} and \eref{defC} and contain sourcing and parametric enhancements from the background solution. As already mentioned in Section \ref{sec:modelsetup}, to numerically solve this ODE, because it falls off relatively slowly, one can use a shooting method that matches to the asymptotical solution of this equation at a large $r$ to improve the accuracy.

\begin{figure}[tbp]
\centering
\includegraphics[width=7cm]{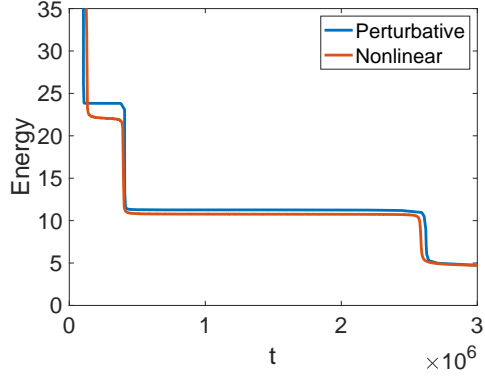}
\caption{Comparison of lifespans computed from the full nonlinear simulations and the leading perturbative analysis. The coupling is chosen to be $g=0.6$ and the initial parameters are chosen to be $A=1$, $a=81$ and $\sigma=90$. }
\label{fig:dedt2omega}	
\end{figure}

The decay rate of the oscillon can be obtained by computing at a large $r$ the $0r$ component of the energy-momentum tensor of the scalar field, $T_{0r}$, which is the energy flux in the $r$ direction. The background field $\Phi_0$ goes like $e^{-\sqrt{1-\omega^2}r}/r^{1/2}$ for large $r$, so its contribution to $T_{0r}$ can be safely neglected. To the leading order approximation, we then have
\begin{equation}
T_{0r}(t,r\to \infty)=\partial_t [A_3^0(r)\cos(3\omega t)]\;\partial_r [A_3^0(r)\cos(3\omega t)]. 
\end{equation}
Asymptotically, $A_3^0(r)$ goes like $A_3^0(r)=A \cos{(kr+\alpha)}/r^{1/2}$ where $k=\sqrt{9\omega^2-1}$ and $A$ and $\ai$ are constants. (The value of $A$ can be obtained by solving \eref{cal:A30} and extracting $A=\max(r^{1/2}A_3^0)$ at large $r$ around $r=60\sim 100$.) So $T_{0r}(t,r)$ in the far field can be further approximated by
\be
T_{0r}(t,r\to \infty)=\partial_t \[\frac{A}{r^{1/2}}\cos(kr+\alpha)\cos(3\omega t)\]\; \partial_r \[\frac{A}{r^{1/2}}\cos(kr+\alpha)\cos(3\omega t)\].
\ee
To compute the decay rate, we only include the outgoing waves of the perturbative field $A^0_3$, which leads to
\be
T^{\rm (out)}_{0r}(t,r\to \infty)=\partial_t \[\frac{A}{2r^{1/2}}\cos(kr-3\omega t+\alpha)\]\;\partial_r \[\frac{A}{2r^{1/2}}\cos(kr-3\omega t+\alpha)\].
\ee
Averaging over a few temporal oscillations and integrating over a circle at a large $r$, the energy decay rate of the oscillon (energy radiated away in a unit time by the oscillon) in the leading perturbative approximation is given by
\be
P(\oi)= 2\pi r \left<T^{(\rm out)}_{0r}(t,r\to \infty)\right>_{t}= \frac{3\pi\omega kA^2}{4} =\f{3\pi\omega k}{4} \[\max(r^{\f12}A_3^0)\]^2,
\label{cal:decayrate}
\ee
where the average $\left<\right>$ is over time $t$. Therefore, for each frequency $\oi$, we can obtain an energy decay rate. 
In terms of the energy decay rate, the reason for the existence of the multiple energy plateaus is because there are multiple frequencies (the frequencies of the levels identified in the last subsection) where the decay rate is exponentially suppressed. Having obtained the energy decay rate, one can then compute the lifespan of an excited oscillon level when it migrates from a lower frequency $\omega_1$ to a higher frequency $\omega_2$ by evaluating
\be
\ti_{12}=\int_{\omega_1}^{\omega_2}\frac{\d E(\omega)/\d \oi}{P(\omega)}\d\omega,
\ee
where $E(\omega)$ is the energy of the background solution for a fixed $\oi$, which can be obtained by integrating \eref{back pde} for various $\oi$. So if we want to evaluate the lifetime of, say, the 2nd energy plateau, we should let $\oi_1<\oi_{\rm 2nd}<\oi_2$, where $\oi_{\rm 2nd}$ is the dominant frequency of the 2nd level. In this numerical evaluation, it is important that we should sample the frequencies close to the plateau frequency such as $\oi_{\rm 2nd}$ very finely, as the decay rates at those frequencies are significantly suppressed, which is precisely the reason to have excited oscillons at those frequencies. In Figure \ref{fig:dedt2omega}, we see that the lifetimes of the excited oscillons computed from this perturbative approach matches those of the previous full nonlinear simulations rather well. 

Lastly, we would like to point out that the excited spherical oscillons we studied in this paper can be intuitively viewed as a node-less background plus a leading multi-node perturbation, in terms of the expansion (\ref{modeexp}). Here, the node does conform to the usual meaning of a stationary point in the field's evolution. This underlies the difference between the energy density profile of the first plateau and those of the higher plateaus in Figure \ref{fig:Hfig}. Solutions with a multi-node background also exist, which are morally closer to the excited Q-balls studied in \cite{hep-th/0205157,1206.2930, 2004.03446, 2112.00657, 2201.09239} and which we have also explicitly constructed, but the lifespans of these solutions are much shorter. For example, for the same setup as Figure \ref{fig:energyevolution}, the solution with one-node background only exists for a time duration of  less than $400$.

\section{Multipolar oscillons}

\label{sec:multipolarOSC}

In the previous section we focused on the oscillons that are spherically symmetric. We have uncovered a tower of excited spherical oscillons that have higher energies than the most quasi-stable oscillon. These excited oscillons have shorter lifespans as their energies increase, and can naturally emerge in succession from the evolution of a dense lump that is initially sufficiently close to these excited oscillons. In this section, we will explore oscillons that are complex in a complementary direction -- oscillons that are anisotropical. That is, we will see that there exist quasi-stable structures in real scalar theories that are multipolar. These are similar to spinning Q-balls \cite{hep-th/0205157, hep-th/0302032, gr-qc/0505143, 0804.1357, 0812.3968, 0904.4802, 0907.2801, 0907.0913, 0909.2505, 1207.3715, 1612.05835, 1909.01950}. One major difference is that, due to the realness of the scalar field in the current case, an multipolar oscillon must contain multiple multipoles, although one multipole can dominate its energy density. Also, we have been unable to find the equivalence of charge-swapping Q-balls \cite{1409.3232, 2101.06988, 2202.08392} in this real scalar case.

\subsection{Dipolar oscillons}

\begin{figure}
	\centering
	\includegraphics[width=7cm]{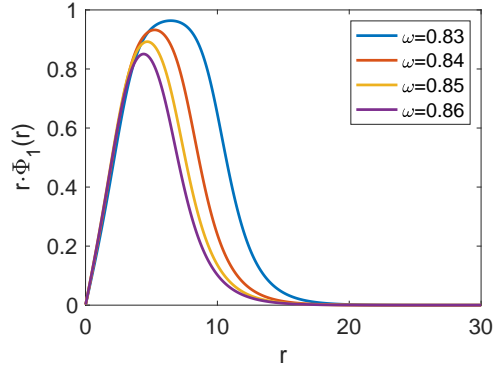}
	\caption{Background profile $\Phi_1(r)$ for different $\omega$. The coupling is chosen as $g=0.64$.}
	\label{fig:backgroundofl0=1}
\end{figure}

In the previous section, it is relatively straightforward to construct excited spherical oscillons, as they seem to be strong attractors that can easily arise from relaxation of some approximate spherical configurations. It is also aided by the absorbing boundary condition which in the spherical case is quite efficient to eliminate the radiation emitted in the relaxation process. The numerical simulations are then checked to a good approximation by the semi-analytical perturbative analysis. Our strategy in this section will however be slightly different, as the multipolar oscillons are less easier to construct. We will prepare the initial configuration of a multipolar oscillon with the perturbative method, although it is sufficient to include only the leading background solution, and then numerically simulate its evolution fully nonlinearly.

\begin{figure}
	\centering
	\includegraphics[width=15.4cm]{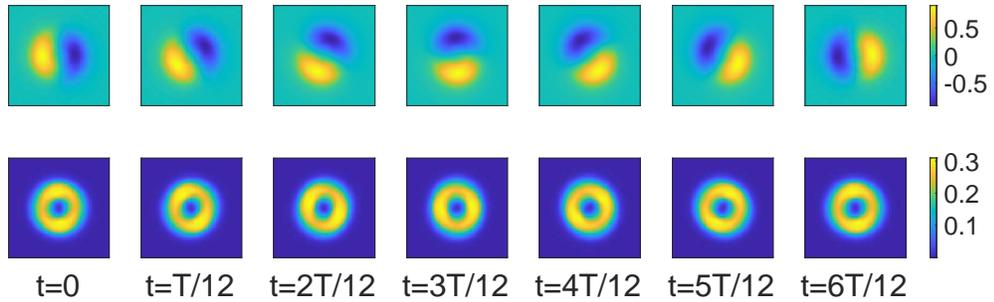}
	\caption{Evolution of the dipolar oscillon. The top row is the evolution of the $\phi$ field, and the bottom is the evolution of the energy density. The rotation period of the field is $T$, while that of the energy density is $T/2$, where $T=2\pi/\oi$ (see \eref{phidefini1} for the definition of $\oi$). Both of them rotate counterclockwise in our initial setup. The coupling is chosen to be $g=0.64$.}
	\label{fig:pic for l1}
\end{figure}

To construct a $l_0$-pole oscillon, we assume that the background field in \eref{modeexp} only have the $l_0$-th multipole and do not have terms with $l<l_0$. For a dipolar oscillon, we have $l_0=1$ and the background field is given by
\be
\label{phidefini1}
\phi(x) = \Phi_1(r) r \cos(\omega t-\theta) +\mc{O}(A) ,
\ee
which is rotating anti-clockwise. We have factored out $r$ such that $\Phi_1$ has $\pd_r\Phi_1(r=0)=0$ as it boundary value at $r=0$. The background field satisfies the following ODE
\begin{equation}
\partial_{r}^2\Phi_1+\frac{3}{r}\partial_r\Phi_1+\omega^2\Phi_1=\Phi_1-\frac{3}{2}r^2(\Phi_1)^3+\frac{15}{8}gr^4(\Phi_1)^5.
\label{back pde l=1}
\end{equation}
The $r\Phi_1$ profiles for a few different $\oi$ are shown in Figure \ref{fig:backgroundofl0=1}. As already anticipated in the factorization of $r\Phi_1$, the background field must vanish linearly in the center for the dipole. Inputting these background multipole solutions as the initial conditions in 2+1D simulations, we will find rotating dipolar solutions. A typical time evolution of the field values of the field and the energy density are shown in Figure \ref{fig:pic for l1}. We see that the field rotates at the same period as the field's dominant oscillation, while the energy density rotates at half of the period of the field's dominant oscillation. The energy density vanishes both in and far away from the center of the oscillon.

 However, we are unable to construct a non-rotating dipolar oscillon. Non-rotating initial setups would only lead to two repelling oscillons, as closely placed oscillating lumps in anti-phase repel each other \cite{1409.3232}. This is different from a complex scalar field with a U(1) symmetry, where it is actually possible to have non-rotating multipolar solutions, dubbed charge-swapping Q-balls \cite{1409.3232}. In that case, due to the luxury of a complex field that has two real components, a Q-ball can be roughly thought of as two oscillons for the two real components respectively, which gives rise to dipolar configurations where the dominant oscillons are actually in phase (for a real scalar, they can only be in anti-phase). This provides attractive forces between the lumps and allows non-rotating dipolar Q-balls to form.

\begin{figure}
	\centering
	\includegraphics[width=9cm]{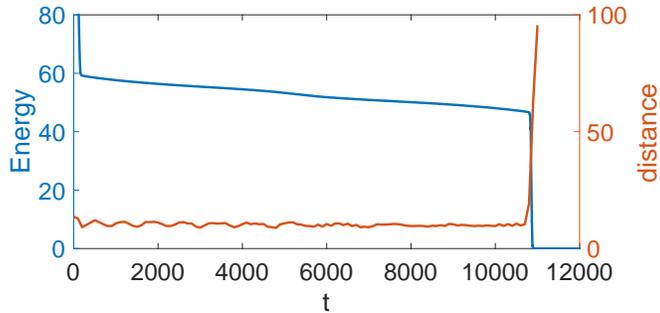}
	\caption{Evolution of the energy and the distance between the dipole peaks for a dipolar oscillon. The dipolar oscillon disintegrates into two lumps flying away from each other at its demise. The parameter choices are the same as those in Figure \ref{fig:pic for l1}.}
	\label{fig:energyevo}
\end{figure}

As perhaps naturally expected, the lifespan of the dipolar oscillon is less than that of the spherical oscillons, with all parameters chosen equal; see Figure \ref{fig:energyevo}. Recall that for the same coupling constant $g$, we have not seen the most stable oscillon decay in our simulations, and, from Figure \ref{fig:numandtofplatform}, we see that even the lifespan of the 2nd or 3rd energy level of the spherical oscillon is significantly longer than that of the dipolar oscillon. Additionally, the energy of a dipolar oscillon is slightly higher than the highest level spherical oscillon we are able to construct. In Figure \ref{fig:energyevo}, we also see that what has been constructed is not a perfect dipole, with the distance between the peak values of field oscillating slightly with time, especially in the first half of its evolution. This is probably partly due to our method of construction which only makes use of the 0-th order approximation with the background field. However, the dipolar oscillon is not a factorizable solution, so some level of multipole mixing is expected. The oscillation does subside in a slow relaxation process. In our simulations, the dipolar oscillon always ends with two lumps flying away from each other, as we can see in Figure \ref{fig:energyevo}.

\begin{figure}
	\centering
	\includegraphics[width=5.5cm]{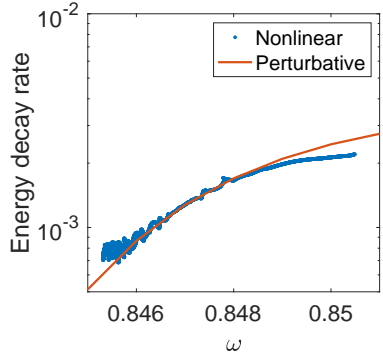}
	\caption{Comparison of the decay rates between the perturbative approach and the nonlinear simulations for the dipolar oscillon for a section of frequencies. The parameter choices are the same as those in Figure \ref{fig:pic for l1}.}
	\label{fig:decayel1}
\end{figure}

We can also utilize the higher order perturbative analysis to estimate the decay rate of the dipolar oscillon, as we did in Section \ref{sec:pertanaly}. Again, since $A^2_2$ vanishes, the lowest order radiation field is $A^3_3$, whose equation of motion is given by
\begin{equation}
\begin{aligned}
\partial_{r}^2A_3^3+\frac{7}{r}\partial_r A_3^3+(9\omega^2-1-B_0^0)A_3^3=\frac{C_3^3}{r^3}+\mc{O}(A^2).
\end{aligned}
\end{equation}
where $C_3^3(r)$ and $B^0_0(r)$ are defined in \eref{defB} and \eref{defC}. Going through similar steps as the spherical case, we can get that the decay rate of the dipolar oscillon is given by
\be
P(\oi)= 2\pi r \left<T^{(\rm out)}_{0r}(t,r\to \infty,\thi)\right>_{t, \theta}=\f{3\pi\omega k}{8} \[\max(r^{\f12}A_3^3)\]^2.
\ee
where here the average $\left<\right>$ is over the time $t$ and the angle $\thi$. We can compare the decay rates from this perturbative approach with those computed from the nonlinear simulations, as shown in Figure \ref{fig:decayel1} for a range of frequencies. This range of frequencies correspond exactly to the evolution of the dipole oscillon in Figure \ref{fig:energyevo} where the distance between the peaks is stabilized.

\begin{figure}
	\centering
	\subfigure[Power spectrum of $\phi(t,r=4.9)$.]{\includegraphics[width=7.5cm]{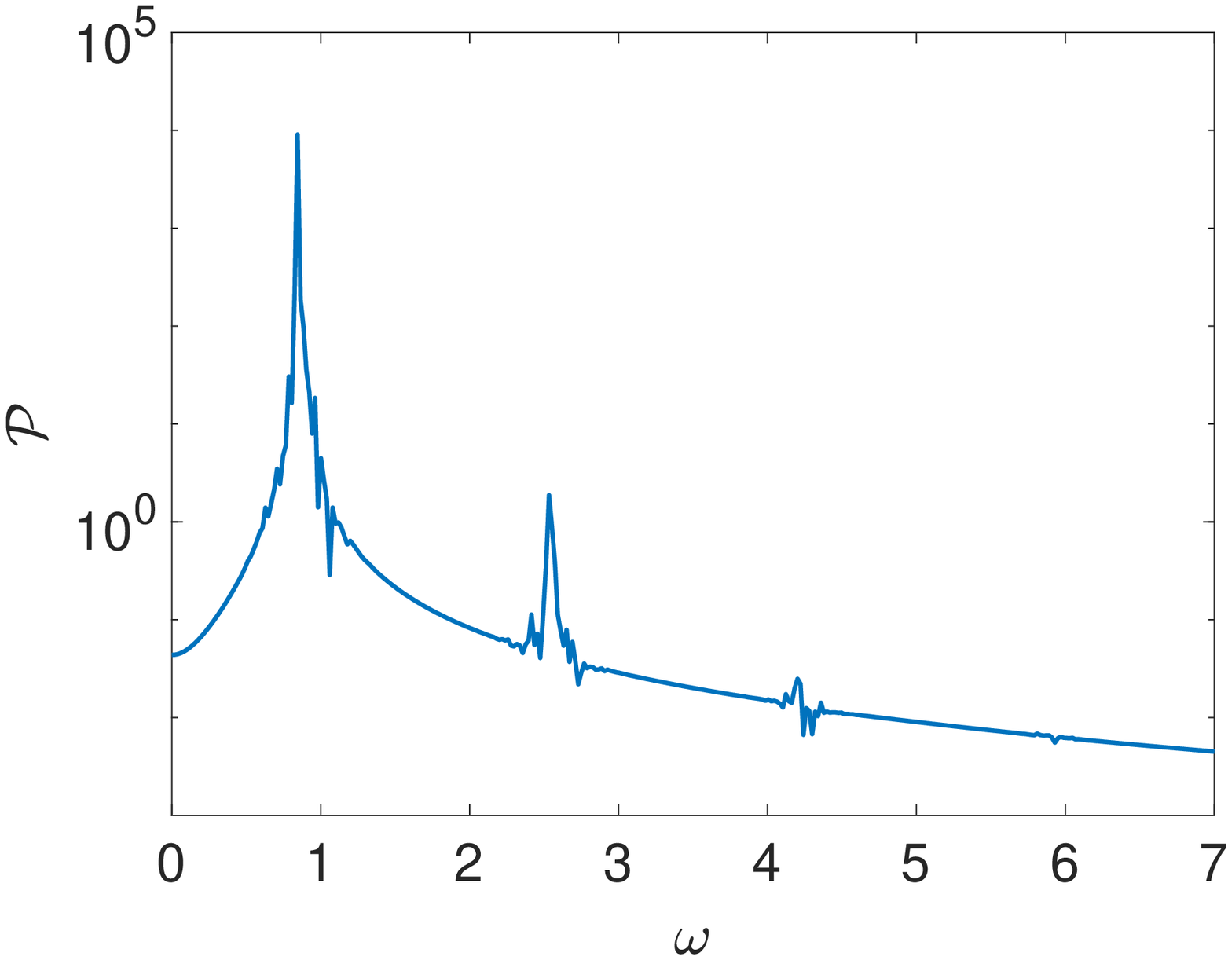}}
	\subfigure[Power spectrum of $\phi(t,r=81.9)$]{\includegraphics[width=7.5cm]{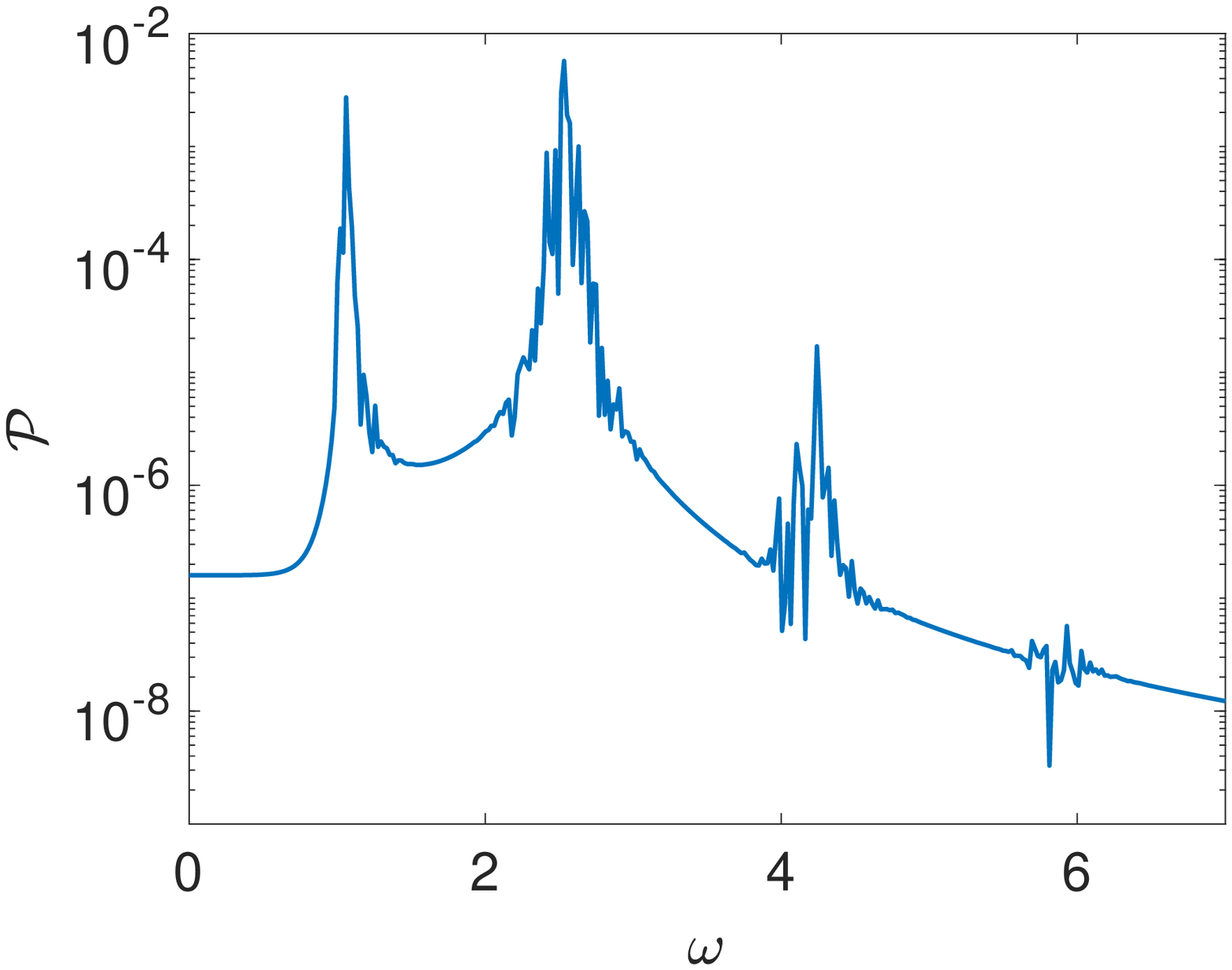}}
	\caption{Fourier Power spectra of the scalar field at two representative points. We can see that  the base frequency dominates the spectrum near the center of the spinning oscillon, while it is the third multiple of the base $\omega$ far away from the center. The parameter choices are the same as those in Figure \ref{fig:pic for l1}. }
	\label{fig:fourierl1}
\end{figure}

It is also instructive to see the oscillation patterns of the field at different places of the dipolar oscillon. In Figure \ref{fig:fourierl1}, we plot the Fourier power spectra of the field at point $r=4.9$ and point $r=81.9$ respectively. (The dipolar oscillon is non-spherical but in 2+1D its profile is spherically symmetric, the difference in the different directions being only in the phase.) We can clearly see that the base frequency dominates the spectrum near the center of the spinning oscillon, while it is the third multiple of the base $\omega$ that is dominating far away from the center. This justifies the approximation we use in the perturbative analysis above using a background field with a base frequency and a radiation field with three multiple of the base.

We would like to stress that the stability of dipolar oscillons is rather sensitive to initial conditions.  This is different from the case of spherical/monopole oscillons, whose ability to form is insensitive to the initial configurations. Therefore, in order to construct a dipolar oscillon, it is essential to solve the equation of motion of the background field (\ref{back pde l=1}) accurately and use it as the initial input. In other words, the ``attractor basin'' of the dipolar oscillon is much smaller than the monopolar oscillon. Also, the end state of a dipolar oscillon is not dissipation of it. For the case of a spherical oscillon, its evolution can be tracked in the frequency space where the oscillon's frequency increases with time until the frequency reaches the upper limit, which is the mass of the field, and then the field oscillation is no longer supported, leading to the oscillon disintegration via dissipation. However, for the dipolar oscillon, the end state is when the attraction between the lumps of the dipole can not support its rotation, and the two lumps fly away from each other. However, both of the two processes happen very quickly.

\subsection{Higher multipole oscillons}

\begin{figure}
	\centering
	\includegraphics[width=15.4cm]{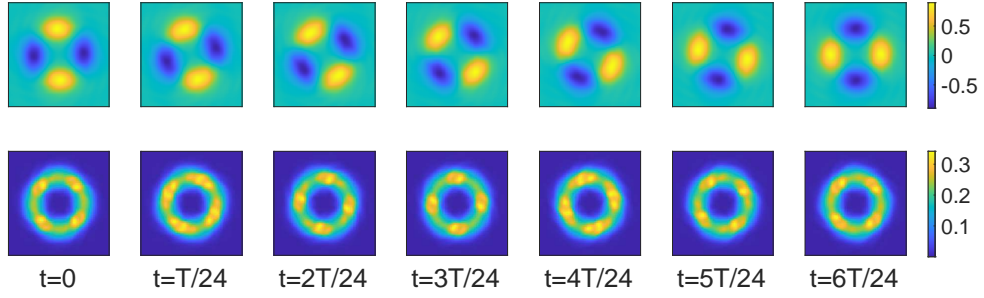}
	\caption{Evolution of the quadrupolar oscillon. The top row is the evolution of the $\phi$ field, and the bottom row is the evolution of the energy density. The rotation period of the $\phi$ field is denoted as $T$, while the rotation period of the energy density is $T/2$, where $T=2\pi/\oi$ (see \eref{phidefini2} for the definition of $\oi$).  The coupling is chosen to be $g=0.64$.}
	\label{fig:pic for l2}
\end{figure}

It is also possible to construct higher multipole oscillons by initially preparing with the leading background field solutions. For example, for the case of $l_0=2$, we only need to consider the following background
\be
\label{phidefini2}
\phi(t,r,\theta)=\Phi_{2}(r) r^{2} \cos(\omega t-2\theta) + \mc{O}(A)  ,
\ee
where $\Phi_{2}$ satisfies the following ODE
\begin{equation}
\partial_{r}^2\Phi_2+\frac{5}{r}\partial_r\Phi_2+\omega^2\Phi_2=\Phi_2-\frac{3}{2}r^4(\Phi^2)^3+\frac{15}{8}gr^8(\Phi_2)^5.
\label{back pde l=2}
\end{equation}
This ODE again can be solved by a shooting method, and, using it as the initial condition, the initial configuration will quickly relax to a quadrupolar oscillon, as shown in Figure \ref{fig:pic for l2}.  Quadrupoles usually have shorter lifetimes than dipoles. Taking $g=0.64$ as an example, dipoles have lifetimes of the scale of $10^4$, while quadrupoles are of the scale of $10^3$. Similar to the dipoles, for the quadrupoles we have observed so far, the final decay process is still disintegration via separation.

\begin{figure}
	\centering
	\includegraphics[width=8cm]{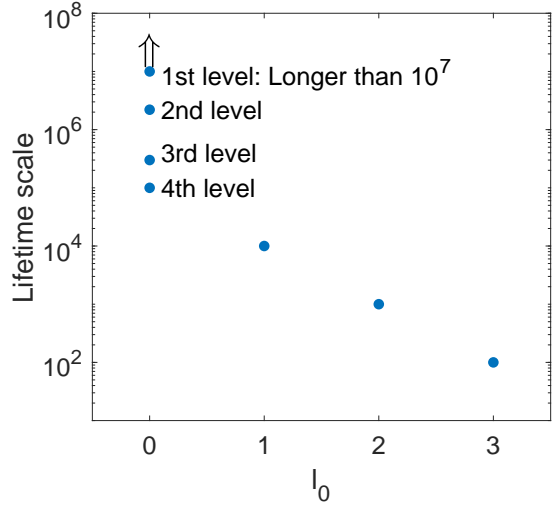}
	\caption{Lifetime scales for leading multipolar oscillons (for coupling $g=0.64$). $l_0$ refers to the multipole of the dominant mode. For the spherical ones ($l_0=0$), we have plotted all the 4 energy levels we found in the last section.}
	\label{fig:lifescale}
\end{figure}

For  even higher multipole oscillons, their lifetimes are much shorter. By again preparing them with the background field method, we find that the sextuple oscillon ($l_0=3$) only lives for a duration of about $10^2$, which is just a dozen of periods of the underlying field oscillations. If we plot the typical lifetime scales of different multipolar oscillons, as shown in Figure \ref{fig:lifescale}, we see that the lifetime scale decreases exponentially with the number of the multipole ($l_0$).  Of course, the lifespans of multipolar oscillons are expected to depend on the potential, as they do for the charge-swapping Q-balls \cite{2101.06988, 2202.08392}. While the above results are quite general in the sense that the polynomial potential we use is the leading approximations in the presence of rapidly converging Taylor expansion, some other potentials such as the logarithmic effective potential commonly arising from loop corrections may lead to excited oscillons with longer lifespans. We leave exploration of this for future work.

\section{Conclusions}
\label{sec:conlusions}

In this paper we have investigated two types of excited oscillons. The first type is spherically symmetric oscillons with higher energies. As its energy increases, such an excited oscillon has an increasingly lower oscillation frequency and shorter lifetime than the un-excited one. They are characterized by increasingly more (approximate) nodes in their radial profiles. We have also preliminarily surveyed the parameter space for their lifetimes numerically. The lifetimes can also be estimated semi-analytically by approximating the oscillon with a factorizable background with a fixed frequency plus radiation fields with higher multiple frequencies. It is often sufficient to only include the leading radiation field. One interesting feature is that, starting from a higher energy oscillon, it will cascade through all the lower energy phases before the oscillon dissipates, and the different phases of the oscillon are connected by rapid bursts of radiation of energy.

The existence of these higher energy spherical oscillons may also be seen by computing the decay rate of the oscillon, by which one finds that there are oscillation frequencies where the rate is exponentially suppressed, leading to these metastable oscillons. Flipping the argument, the underlying reason for the existence of the suppressed decay rates is that there are approximate multi-node oscillon solutions to the equations of motion. The cascading feature of its evolution might also be expected, as the evolution of the oscillon is basically a process where the dominating frequency of the oscillon slowly increases to higher values until reaching the upper limit when the oscillon quickly disintegrates. 

The second type of metastable oscillons we have uncovered are the ones with higher multipoles. Non-spherical structures for Q-balls, cousins of oscillons, have been investigated previously. Similar to the un-excited oscillon, these multipolar oscillons are also not factorizable, and there are necessarily many multipoles involved. Nevertheless, typically, one multipole will dominate, making the overall behavior similar to a multipolar Q-ball. As the multipolar oscillons are more sensitive to the initial conditions than the spherical ones, it is essential that one prepares the initial configuration by the semi-analytical method as so to numerically construct these oscillons. 

The multipolar oscillons we have obtained are the counterparts of the spinning Q-balls \cite{hep-th/0205157}, in which the field profile rotates in one direction. However, we are unable to construct the counterparts of the charge-swapping Q-balls \cite{1409.3232}. The difference again originates from the fact that for oscillons we have only a single component scalar field compared with a complex field for Q-balls, so a multipolar oscillon necessarily has multiple multipoles. For such configurations, aligning the opposite rotating configurations can not be achieved for all the multipoles simultaneously.

The lifespans of the multipolar oscillons are significantly shorter than those spherical excited states. Particularly, for our polynomial model in 2+1D, we have only found reasonably long-lived
oscillons up to the sextuple ones ($l_0=3$; the multipole number $l_0$ here referring to the dominating mode), which only live for tens of oscillations. However, multipolar oscillons may be obtainable for other potentials such as logarithmic ones, as we have seen very long-lived charge swapping Q-balls in a logarithmic potential \cite{2202.08392}.

We have focussed on excited oscillons in the simple model with a renormalizable real scalar field in 2+1D. Generalization of the spherically symmetric oscillons to 3+1D is simple and only involves changing a number in the spherical equation of motion. Indeed, in our preliminary explorations, we have also found the cascading levels in the evolution of the oscillon's energy in 3+1D.  Generalization of the multipolar oscillons to 3+1D is also conceptually straightforward but technically much more involved, as there will be two angle dependence to handle. On the flip side, one would expect much richer phenomenology in higher dimensions. Furthermore, one might also consider models with more complex potentials or/and more scalars. We leave the detailed investigations of these directions to future work.

\acknowledgments

We would like to thank Xiao-Xiao Kou, Anzhuoer Li and Paul Saffin for helpful discussions. SYZ acknowledges support from the starting grants from University of Science and Technology of China under grant No.~KY2030000089 and GG2030040375, and is also supported by National Natural Science Foundation of China under grant No.~11947301, 12075233 and 12047502, and supported by the Fundamental Research Funds for the Central Universities under grant No.~WK2030000036.  \\
~\\

\appendix

\noindent{\bf\Large Appendices}

\section{Numerical codes}
\label{sec:Scode}

As mentioned, we make use of two separate codes to run the nonlinear simulations in this paper, one being a spherical grid code and the other being a Cartesian code. Both of them are endowed with absorbing boundary conditions, as we are interested in the lifetimes of the oscillons.

{\bf Spherical code}:
For the spherical code, symmetrical boundary conditions are introduced at the origin and absorbing boundary conditions in the far field. For our purposes, it is sufficient to use a grid with radius $R=128$, and to choose the radial step size to be $dr=0.25$ and the temporal step size to be $dt=0.01dr$. We use the Runge-Kutta fourth-order method to evolve it temporally. The Courant-Friedrichs-Lewy (CFL) factor is chosen to be relatively small so as to improve convergence, but it is still at least two orders of magnitude faster than the equivalent Cartesian code. 

For the spherical code, the absorbing boundary condition can be implemented as follows. First, note that the equation of motion in $d+1$ dimensions in the far field (large $r$) is given by 
\be
\frac{\partial^2 \phi}{\partial t^2}-\frac{\partial^2 \phi}{\partial r^2}-\frac{d-1}{r}\frac{\partial\phi}{\partial r}+\phi = \mc{O}(\phi^2),
\ee
so the out-going wave solution with frequency $\oi$ is given by  $\phi\simeq \exp[i(\sqrt{\omega^2-1}r-\omega t)]/r^{(d-1)/2}$ at large $r$. For waves whose frequencies are much greater than 1, it can be approximated by $\phi\simeq \exp[i(\omega r-\f{1}{2\oi}r-\omega t)]/r^{(d-1)/2}$. Such out-going waves are solutions of the differential equation
\be
\(\partial_t\partial_r+\partial^2_t+\frac{1}{2}+\frac{d-1}{2r}\partial_t\)\phi=0.
\ee
while the in-going waves are not. Therefore, we can implement this as the boundary conditions at large $r$, which will efficiently absorb out-going waves whose frequencies are larger than the mass of the field.

{\bf Cartesian code}:
The Cartesian code is used to cross-check with the spherical code in the study of cascading oscillons and to run simulations for multipolar oscillons. The code is parallelized in the framework of {\tt LATfield2} \cite{David:2015eya}. The spatial derivatives in the equation of motion are approximated by 4th order finite differences and the time evolution is performed using the Runge-Kutta 4th order method. We make use of Higdon's absorbing boundary conditions \cite{higdon1994radiation,higdon1986absorbing} to absorb out-going waves and reduce unphysical wave reflections at the boundary. These conditions can absorb waves with multiple fixed phase velocities completely and also absorb waves with  a broad range of mismatched phase velocities rather well, provided the parameters of the fixed phase velocities are chosen judiciously. A 2nd order Higdon's condition implemented at the boundaries in the $x^i$ direction is 
\be 
\( \f{\p^2}{\p t^2} \pm (c_1+c_2) \f{\p}{\p x^i \p t} + c_1c_2 \f{\p^2}{\p (x^i)^2} \) \phi ~ \bigg |_{x^i=a} =0.
\ee 
where $+(-)$ indicates the condition to be applied at the upper (lower) $x^i$ boundary and $a$ is the location of the boundary. $c_j$'s are tunable constants, which should be chosen to be close to the phase velocities of the far field waves. In our case, we simply set $c_1=c_2=1$. See \cite{2101.06988} for more details. For the other numerical setup, the spatial grid is chosen to have $1024 \times 1024$ sites and the physical spacing between adjacent sites is $dx=0.2$. The time step is chosen to be $dt=0.02$ to satisfy the CFL condition.

\section{Decoupling between different sets of $A^l_n$ modes}
\label{sec:decouple}

In this Appendix, we shall analyze the structure of the perturbative equations of motion and separate out the perturbative modes that are sourced by the background from those that are decoupled from the background. These decoupled modes include the $A^l_n$ modes with negative $l$.

Consider a counter clockwise spinning background field $\Phi_{l_0}$ ($l_0>0$) supplemented with both positive and negative $l$ perturbative field $A_n^l$
\begin{equation}
\phi(t,r,\theta)=\Phi_{l_0}(r) r^{l_0} \cos(\omega t-l_0\theta) + \sum_{n>1,|l|>l_0}A_n^l(r) r^{|l|} \cos(n\omega t-l\theta).	
\end{equation}
The perturbation equation of motion takes the following form:
\be
\partial_{r}^2A_n^l+\frac{2|l|+1}{r}\partial_r A_n^l+(n^2\omega^2-1-B^0_0)A_n^l -\frac{C_n^l}{r^{|l|}}
=\frac{1}{2r^{|l|}}\sum_{j=2}^{\infty}B_j^{jl_0}\(r^{|l+jl_0|}A_{n+j}^{l+jl_0}+r^{|l-jl_0|}A_{n-j}^{l-jl_0}\) ,
\ee
where $C_n^l$ and $B_n^l$ are defined by \eref{defC} and \eref{defB} respectively. Re-labeling $l$ as $l=nl_0+k$, where $k$ is an integer with $|nl_0+k|>l_0$, we can re-write the above equation as
\bal
\partial_{r}^2A_n^{nl_0+k}+ &\frac{2|nl_0+k|+1}{r}\partial_r A_n^{nl_0+k}  +(n^2\omega^2-1-B^0_0)A_n^{nl_0+k} -\frac{C_n^{nl_0+k}}{r^{|nl_0+k|}}
\\
&=\frac{1}{2r^{|nl_0+k|}}\sum_{j=2}^{\infty}B_j^{jl_0}\(r^{|(n+j)l_0+k|}A_{n+j}^{(n+j)l_0+k}+r^{|(n-j)l_0+k|}A_{n-j}^{(n-j)l_0+k}\) .
\eal 
An immediate observation is that the above set of equations separate into a number of subsets of equations that are labeled by $k$, with the subsets decoupled from each other. In other words, the $A^{nl_0+k}_n$ modes with the same $k$ are coupled, while the $A^{nl_0+k}_n$ modes with different $k$ are decoupled from each other.  Also,  $C_n^{nl_0+k}$ is nonzero only when $k=0$. That is, all the $A^{nl_0+k}_n$ modes with nonzero $k$ are unsourced -- the $A^{nl_0+k}_n$ with the same nonzero $k$ satisfied a system of unsourced/homogeneous linear ODE, and therefore these modes  are not excited by our background field $\Phi_{l_0}$. These unsourced modes include the modes where $A_n^l$ with $l<0$.

\section{Convergence study of cascading oscillons}
\label{sec:sph3Dsim}

In Section \ref{sec:casosci}, we made use of a spherical code to simulate cascading oscillons, which speeds up exploration of the parameter space. The spherical code also allows implementation of more efficient absorbing boundary conditions, but has the disadvantage of neglecting non-spherical modes, which may contribute to the decay of the oscillons. The generic expectation is that for a spherical oscillon the dominated decay mode is the monopole mode. In this section, we shall verify that this is indeed the case by comparing a typical evolution in the spherical simulation with the corresponding 2+1D simulations.

\begin{figure}
	\centering
	\includegraphics[width=8cm]{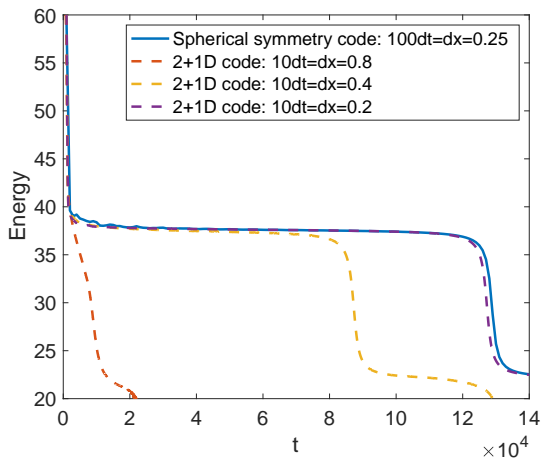}
	\caption{Validation of the spherical code with 2+1D simulations. The spherical code is used to map the parameter spaces of cascading oscillons. The parameters are chosen as $g=0.60$.}
	\label{conver}
\end{figure}

Figure \ref{conver} shows an evolution of a typical cascading oscillon. We see that the lifetimes of the levels of the oscillons in the 2+1D code converge to that in the spherical code as the accuracy increases. Quantities such as lifetimes is more difficult to determine numerically than quantities that can be measured at one time instance, as numerical errors can accumulate with time in the former case. Figure \ref{conver} highlights that the accuracy in the 2+1D code needs to be higher than that in the spherical code to converge to the actual result. For a comparison of computing resources needed for the matched runs in Figure \ref{conver} (the spherical code and the most accurate 2+1D code), the 2+1D code uses a grid of $1024\times1024$ points and requires more than $200$ wall-clock hours using 16 threads to reach $t=1.4\times 10^5$, while the spherical symmetric code only needs a $256\times 2$ grid and takes about 3 wall-clock hours using 16 threads to reach the same physical time, which is the reason we used the spherical code to perform the parameter space survey. This justifies the use of the spherical code to analyze the properties of the spherical oscillons.

\bibliographystyle{JHEP}
\bibliography{refs}

\end{document}